\documentclass[manuscript]{acmart}

\AtBeginDocument{%
  }
\usepackage{graphicx}
\usepackage{textcomp}
\usepackage{algorithmic}
\usepackage[ruled,linesnumbered]{algorithm2e}
\usepackage{xcolor}
\usepackage{xspace}
\usepackage{tcolorbox}
\usepackage{booktabs,multirow,stfloats}
\usepackage{fontawesome}
\usepackage{fbox}
\usepackage{pifont}
\usepackage{xspace}
\newcommand{\approach}{{\sc PATCH}\xspace}

\definecolor{light-green}{RGB}{227,242,217}
\definecolor{light-yellow}{RGB}{255,242,202}
\definecolor{light-red}{RGB}{250,219,223}
\definecolor{light-blue}{RGB}{218,227,245}
\definecolor{light-blue2}{RGB}{72,116,203}
\definecolor{light-blue3}{RGB}{0,255,255}
\definecolor{light-gray}{RGB}{217,217,217}

\setcopyright{acmlicensed}
\copyrightyear{2024}
\acmYear{2024}
\acmDOI{XXXXXXX.XXXXXXX}

\acmJournal{TOSEM}
\acmVolume{37}
\acmNumber{4}
\acmArticle{111}
\acmMonth{8}

\begin{document}

\title[Empowering Large Language Model with \approach for Automatic Bug Fixing]{\approach: Empowering Large Language Model with Programmer-Intent Guidance and Collaborative-Behavior Simulation for Automatic Bug Fixing}

\author{Yuwei Zhang}
\email{zhangyuwei@iscas.ac.cn}
\orcid{0009-0008-1016-7361}
\affiliation{%
  \institution{Key Laboratory of System Software (Chinese Academy of Sciences), Institute of Software, Chinese Academy of Sciences; University of Chinese Academy of Sciences}
  \city{Beijing}
  \country{China}
}

\author{Zhi Jin}
\authornote{Corresponding authors}
\email{zhijin@pku.edu.cn}
\orcid{0000-0003-1087-226X}
\affiliation{%
  \institution{Key Laboratory of High Confidence Software Technologies (Peking University), Ministry of Education; School of Computer Science, Peking University;}
  \city{Beijing}
  \institution{School of Computer Science, Wuhan University}
  \city{Wuhan}
  \country{China}
}

\author{Ying Xing}
\email{xingying@bupt.edu.cn}
\orcid{0000-0003-2807-1911}
\affiliation{%
  \institution{School of Intelligent Engineering and Automation, Beijing University of Posts and Telecommunications}
  \city{Beijing}
  \country{China}
}

\author{Ge Li}
\authornotemark[1]
\email{lige@pku.edu.cn}
\orcid{0000-0002-5828-0186}
\affiliation{%
  \institution{Key Laboratory of High Confidence Software Technologies (Peking University), Ministry of Education; School of Computer Science, Peking University}
  \city{Beijing}
  \country{China}
}

\author{Fang Liu}
\email{fangliu@buaa.edu.cn}
\orcid{0000-0002-3905-8133}
\affiliation{%
  \institution{State Key Laboratory of Complex \& Critical Software Environment; School of Computer Science and Engineering, Beihang University}
  \city{Beijing}
  \country{China}
}

\author{Jiaxin Zhu}
\authornote{Affiliated with Nanjing Institute of Software Technology, University of Chinese Academy of Sciences, Nanjing, China.}
\email{zhujiaxin@otcaix.iscas.ac.cn}
\orcid{0000-0002-0905-2355}
\author{Wensheng Dou}
\authornotemark[2]
\email{wsdou@otcaix.iscas.ac.cn}
\orcid{0000-0002-3323-0449}
\author{Jun Wei}
\authornotemark[1]
\authornotemark[2]
\email{wj@otcaix.iscas.ac.cn}
\orcid{0000-0002-8561-2481}
\affiliation{%
  \institution{Key Laboratory of System Software (Chinese Academy of Sciences), Institute of Software, Chinese Academy of Sciences; University of Chinese Academy of Sciences}
  \city{Beijing}
  \country{China}
}

\renewcommand{\shortauthors}{Zhang et al.}

\begin{abstract}
  Bug fixing holds significant importance in software development and maintenance. Recent research has made substantial strides in exploring the potential of large language models (LLMs) for automatically resolving software bugs. However, a noticeable gap in existing approaches lies in the oversight of collaborative facets intrinsic to bug resolution, treating the process as a single-stage endeavor. Moreover, most approaches solely take the buggy code snippet as input for LLMs during the patch generation stage. To mitigate the aforementioned limitations, we introduce a novel stage-wise framework named \approach. Specifically, we first augment the buggy code snippet with corresponding dependence context and intent information to better guide LLMs in generating the correct candidate patches. Additionally, by taking inspiration from bug management practices, we decompose the bug-fixing task into four distinct stages: bug reporting, bug diagnosis, patch generation, and patch verification. These stages are performed interactively by LLMs, aiming to simulate the collaborative behavior of programmers during the resolution of software bugs. By harnessing these collective contributions, \approach effectively enhances the bug-fixing capability of LLMs. We implement \approach by employing the powerful dialogue-based LLM ChatGPT. Our evaluation on the widely used bug-fixing benchmark BFP demonstrates that \approach has achieved better performance than state-of-the-art LLMs.
\end{abstract}



\received{20 February 2007}
\received[revised]{12 March 2009}
\received[accepted]{5 June 2009}

\maketitle

\section{Introduction}
\label{int}

\begin{figure}[tbp]
  \centering
  \includegraphics[width=0.99\textwidth]{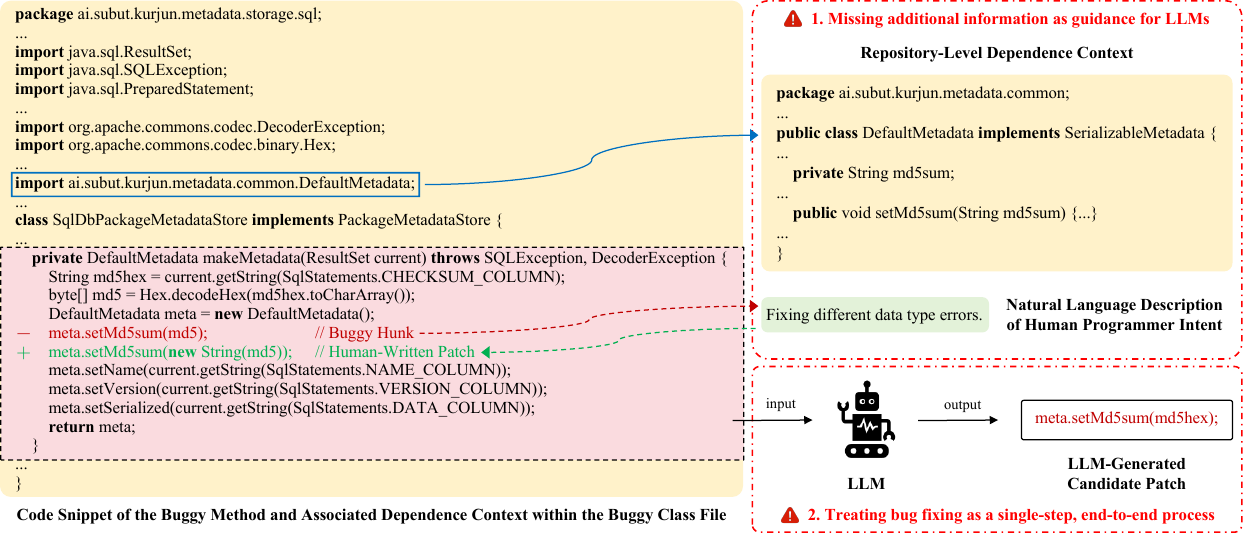}
  \caption{Limitations of Existing LLM-Based Bug-Fixing Approaches.}
  \label{limitation}
  \Description{}
\end{figure}

Software systems, by virtue of inherent complexity and inadequate testing, inevitably contain bugs that can lead to substantial losses \cite{yuan2014simple,wong2017familiar}. To expedite the resolution of software bugs, automatic bug fixing \cite{zhang2016literature} has been proposed as a means to mitigate the costs associated with software debugging. Traditional approaches generally entail mutating the buggy code through predefined search strategies \cite{goues2012genprog,ghanbari2019practical}. Nevertheless, these approaches face challenges primarily due to the time-intensive nature of validation strategies, such as verifying the correctness of generated patches using exhaustive test suites. With the rapid advancements in deep learning (DL), there has been a surge of interest in neural-based bug-fixing approaches \cite{zhong2022neural,zhang2024survey}, exploiting the powerful representation capabilities of DL models to autonomously learn bug-fixing patterns. However, existing neural-based approaches \cite{lutellier2020coconut,jiang2021cure,zhu2021syntax,ye2022neural,zhang2023neural,zhu2023tare} collect historical bug-fixing datasets sourced from open-source code repositories for supervised training, which may restrict their generalizability to unseen bug types \cite{xia2022less}. More recently, researchers have commenced leveraging large language models (LLMs) for bug fixing without the necessity of fine-tuning. The application of LLMs to bug fixing \cite{xia2023automated,jiang2023impact,huang2023empirical} involves devising prompts that can consist of either the buggy code alone or a combination of the buggy code and a few task-specific bug-fixing pairs, with the goal for LLMs to learn from the provided prompts and generate patches for the given buggy code. While current LLM-based approaches have demonstrated promising results compared to previous neural-based techniques, they still possess certain limitations as illustrated in Figure~\ref{limitation}.

\textbf{Limitation 1: Missing additional information associated with the buggy code as guidance for LLMs.} The left part of Figure~\ref{limitation} presents a motivating example derived from the real-world bug-fixing benchmark BFP \cite{tufano2019empirical}, which comprises the buggy method, highlighting its fault location (i.e., the buggy hunk), alongside the ground-truth patch written by the corresponding programmer. When provided with insufficient input information, the LLM outputs an incorrect candidate patch, predicting a different token (i.e., the variable \textbf{\texttt{md5hex}}) based on the buggy method content to replace the input parameter (i.e., the variable \textbf{\texttt{md5}}) of the function \textbf{\texttt{setMd5sum}} invoked via the variable \textbf{\texttt{meta}} within the buggy hunk. Notably, even experienced programmers find it challenging to identify the appropriate patch for fixing the buggy hunk by examining the code snippet of the buggy method alone. In this case, the object \textbf{\texttt{DefaultMetadata}} initialized by \textbf{\texttt{meta}} represents a cross-file data dependency within the corresponding code repository, which is evident from the import information (framed by the {\color[RGB]{0,112,192}blue rectangle}) provided in the buggy class file. Intuitively, when the repository-level dependencies (i.e., the contextual information of \textbf{\texttt{setMd5sum}}) are utilized as prompt inputs, the LLM can deduce that the data type of \textbf{\texttt{md5}} (i.e., \textbf{byte[]}) does not conform to the required input parameter type of \textbf{\texttt{setMd5sum}} (i.e., \textbf{String}). Furthermore, when human programmers encounter a bug during the real-world software development environment, they begin by debugging the buggy hunk, considering its surrounding context, and analyzing compiler-generated error messages to identify the root cause of the bug. Subsequently, they document their intent by summarizing the key information required to fix the corresponding bug via natural language. By considering the intent description provided by the human programmer (i.e., \textit{fixing different data type errors}), we hypothesize that the LLM can reason more effectively about the necessary modifications for patching the buggy hunk.

\textbf{Limitation 2: Treating the task of bug fixing as a single-step, end-to-end procedure.} In practice, bug fixing is a multifaceted task wherein every discovered bug undergoes a specific and intricate process before being effectively resolved \cite{eren2023analyzing}. Although LLMs have demonstrated capabilities akin to human logical understanding \cite{wei2022chain}, they continue to struggle with resolving bugs that involve nested program structures and cross-file dependence relationships. Software debugging inherently necessitates multi-step reasoning, a process that poses a considerable obstacle for LLMs that predominantly rely on pattern recognition rather than authentic cognitive processes. Consequently, this reliance limits the effectiveness of LLMs in resolving complex bugs. Human programmers, by contrast, tend to seek teamwork as a means of tackling intricate debugging-related tasks in software engineering (SE) practices \cite{mcChesney2004communication,lindsjorn2016teamwork}. However, current LLM-based bug-fixing approaches typically focus on directly utilizing LLMs to generate candidate patches under the setting of zero-shot or few-shot prompting paradigms, neglecting the interactive and collaborative behaviors exhibited by human programmers (e.g., one reviewer will be responsible for verifying the correctness of the candidate patch generated by the corresponding developer) during the resolution of complex software bugs. Zeng et al. \cite{zeng2022extensive} conducted a comprehensive evaluation of bug-fixing performance using eight open-access state-of-the-art LLMs on the BFP benchmark. Their empirical findings reveal that all the evaluated LLMs exhibit low accuracy, with performance rates below 15\% in the task of bug fixing. Given that the effectiveness of LLMs is highly contingent on the surface structure of the prompts used \cite{zhao2021calibrate}, it is imperative to explore more effective prompting techniques to enhance the ability of LLMs in generating correct patches.

\textbf{To bridge the gap between the capabilities of LLMs and human programmers in bug fixing, this paper presents a stage-wise framework, referred to as} \approach. This framework introduces two novel mechanisms that empower the LLM with \textbf{P}rogr\textbf{A}mmer-in\textbf{T}ent guidance and \textbf{C}ollaborative-be\textbf{H}avior simulation. These mechanisms effectively address the two aforementioned limitations, resulting in a significant improvement in the bug-fixing performance of LLMs. The specific details of \approach are outlined as follows.

\textbf{Novelty 1: Augmenting the buggy code snippet with additional dependence context and guided programmer's intent as input to the LLM.} This paper employs the widely-used benchmark BFP \cite{tufano2019empirical} for evaluation, which includes an extensive collection of paired bug-fixing instances across a diverse range of real-world bugs, rather than limiting the scope to specific bug types \cite{mastropaolo2024rise}. We commence by enhancing the buggy code snippet through the incorporation of code dependency contexts extracted from both the class and repository levels. Additionally, we integrate the programmer's intent, articulated in natural language, as supplementary guidance. \textbf{Effective bug fixing necessitates a comprehensive grasp of both code semantics and the intent of human programmers. This augmentation aims to guide LLMs in generating effective and accurate candidate patches.} It is vital to leverage sufficient contextual information as fixing ingredients for patch generation. Moreover, understanding the context of a buggy code snippet is essential for human programmers in identifying the root cause of the discovered bug and proposing potential bug-fixing suggestions. Therefore, collecting adequate contextual information from the corresponding GitHub repository can be more effective in generating the correct patches. Furthermore, given the lack of test suites for the BFP benchmark, we leverage commit messages, which are written in natural language, as proxies for the programmer's intent. These messages serve as crucial artifacts within the continuous software development process \cite{xu2019commit}. As illustrated in Figure~\ref{limitation}, it is important to note that the information provided solely describes the type of bug (e.g., \textit{fixing different data type errors}) without detailing the specific process required for its resolution. Consequently, this information can be utilized to accurately simulate the actual development process, minimizing the risk of information leakage issue that could lead to unintended advantages for LLMs. Recent empirical investigations \cite{chakraborty2021multi} have shown that leveraging the commit messages authored by corresponding programmers as supplementary guidance enhances the performance of LLMs by narrowing down the search space. Thus, we collect the bug-fixing commits associated with the buggy code snippets from GitHub to automatically simulate the programmer's intent. Such information facilitates LLMs in better comprehending the desired fixing goals, thereby improving the generation of candidate patches.

\begin{figure}[tbp]
  \centering
  \includegraphics[width=0.71\textwidth]{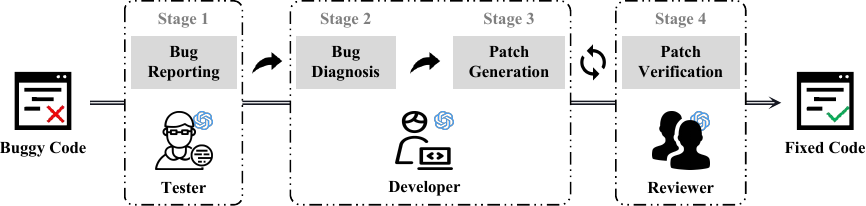}
  \caption{The Brief Structure of \approach.}
  \Description{}
  \label{brief}
\end{figure}

\textbf{Novelty 2: Empowering the bug-fixing performance of LLMs by simulating the collaborative behaviors of programmers via effective bug management practices.} While LLMs trained on code have achieved commendable efficacy in aiding human programmers, their proficiency diminishes notably when confronted with complex debugging-related SE tasks that require multi-step logical reasoning within programs. More recently, researchers have demonstrated the impressive capabilities of LLMs in generating helpful outcomes when tasks are disassembled into a set of modular units with precise queries \cite{wei2022chain,merow2023ai}. Recognizing the significance of an efficient bug management process for successful bug fixing \cite{ohira2012impact}, we decompose the task of bug fixing into four distinct stages: bug reporting, bug diagnosis, patch generation, and patch verification. \textbf{Drawing inspiration from bug management practices, we closely examine the programmers involved at various stages of the bug's life cycle and analyze the impact of their interactions on improving the efficiency of bug fixing.} As depicted in Figure~\ref{brief}, \approach aims to imitate the collaborative problem-solving abilities exhibited by programmers (i.e., the tester, the developer, and the reviewer) throughout the entire life cycle of a bug. To be specific, effective bug resolution initially relies on the tester's comprehensive understanding of the bug, leading to the filing of a detailed bug report. This report provides essential information to the developer for successfully resolving the bug. Within \approach, the developer has a two-fold responsibility. Firstly, the developer engages in the diagnosis process by consulting historical bug corpus and conducting self-debugging. Secondly, the developer generates the candidate patch, guided by the information obtained in the previous stages. Since the correctness of the candidate patch generated on the first attempt cannot be guaranteed, the reviewer's involvement in \approach becomes crucial. The reviewer provides the feedback and collaborates with the developer throughout the workflow, playing a vital role in ensuring the correctness of the generated patch.

In summary, \approach breaks down the bug-fixing task into smaller, more manageable subtasks with the aim of improving the accuracy of automatic bug fixing by incorporating additional information and implementing efficient bug management practices. Moreover, by involving multiple programmers, \approach can enable the inclusion of diverse perspectives and feedback to facilitate the bug-fixing process, thereby mitigating misunderstandings and ensuring the quality of the generated candidate patches. Given the remarkable advancements in generative artificial intelligence (AI), LLMs (e.g., ChatGPT \cite{openai2022chatgpt}) have exhibited commendable performance across various SE tasks \cite{niu2022deep,ozkaya2023application,bano2024large}, opening avenues for inter-model interaction and collaboration. Specifically, \approach employs three ChatGPT agents, each playing a distinct programmer role as showcased in Figure~\ref{brief}, to emulate collaborative efforts in real-world bug management practices. The main contributions of this paper can be summarized as follows:
\begin{itemize}
  \item We present the first attempt at enhancing the capabilities of LLMs for automatic bug fixing by leveraging effective bug management practices. Our alignment approach simulates the interactive behavior of programmers engaged in bug management, which enables LLMs to collaborate and generate correct patches.
  \item We introduce a stage-wise framework called \approach, consisting of three ChatGPT agents, each responsible for specific stages within the bug management process via system instructions and prompts. Our proposed framework shifts the focus from end-to-end bug fixing to a conversation-driven text-generation task, enhancing the understanding and utilization of LLMs in bug fixing.
  \item We construct a meta-rich bug-fixing benchmark that incorporates additional dependence context and the programmer's intent associated with the buggy code snippet. This augmentation aims to provide better guidance to LLMs in generating the correct patches for complex bugs.
  \item We conduct extensive experiments on publicly available bug-fixing benchmarks and thoroughly evaluate each component of the proposed framework. The experimental results demonstrate that \approach surpasses state-of-the-art LLMs, highlighting its superior performance.
  \item We publicly release the replicate package \cite{zhang2024patch} of \approach on Zenodo. The open-sourced artifacts can better support the researchers in the SE community in reproducing \approach.
\end{itemize}

\textit{Article Organization.} The remainder of this paper is organized as follows: Section~\ref{met} introduces in detail the proposed framework. Section~\ref{exp} and Section~\ref{res} provide the experimental setups and results of our research. Section~\ref{dis} presents case studies and discloses the threats to the validity of our approach. Section~\ref{rel} describes the related work. Section~\ref{con} draws conclusions and indicates directions for future work.

\section{The Proposed Framework \approach}
\label{met}

\begin{figure}[htbp]
  \centering
  \includegraphics[width=\textwidth]{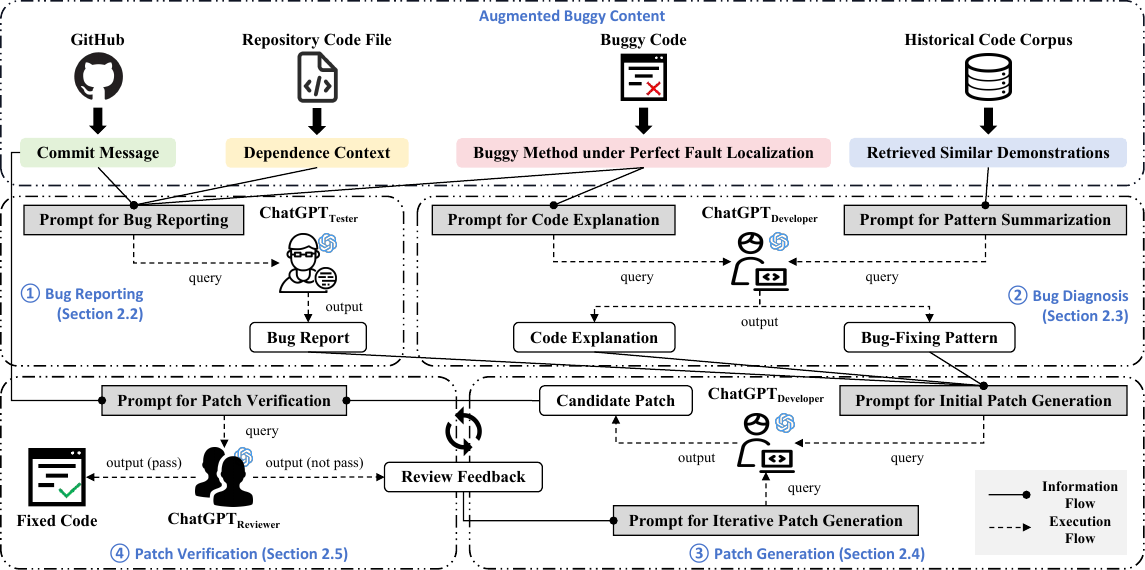}
  \caption{Overview of \approach.}
  \label{overview}
  \Description{}
\end{figure}

In order to mitigate the limitations discussed in Section~\ref{int} concerning existing approaches, we present a novel stage-wise framework dubbed \approach. This framework embodies a programmer-like behavior simulation aimed at augmenting LLMs in the task of bug fixing, leveraging beneficial SE practices (i.e., bug management). Within the process of bug management practice, human programmers engage in collaborative efforts to discover, report, and resolve software bugs, constituting a pivotal facet of the software development life cycle. Such standardized practices not only enhance collaboration efficacy within development teams but also serve to uphold software quality. With this inspiration, our main objective in this paper is to design system components that emulate the cognitive processes of different programmers involved in the bug management process by utilizing the powerful conversation-based ChatGPT model. As illustrated in Figure~\ref{overview}, \approach involves three ChatGPT agents (i.e., $\mathbf{ChatGPT_{Tester}}$, $\mathbf{ChatGPT_{Developer}}$, and $\mathbf{ChatGPT_{Reviewer}}$), each assigned to specific stages (i.e., {\color{light-blue2}\textbf{Bug Reporting}}, {\color{light-blue2}\textbf{Bug Diagnosis}}, {\color{light-blue2}\textbf{Patch Generation}}, and {\color{light-blue2}\textbf{Patch Verification}}) within the bug management process.

\subsection{Task Formulation}
\label{tafo}

Before delving into the details of \approach, this subsection provides a formal description of the bug-fixing task. Following recent studies that utilize LLMs for bug fixing \cite{prenner2022openai,sobania2023analysis}, this paper focuses on fixing single-hunk bugs written in the Java programming language. In this scenario, the generation of candidate patches necessitates program modifications, such as deletions, insertions, or replacements, either at a single line or within a consecutive chunk of code. This process operates under the assumption of perfect fault localization, meaning that LLMs are aware of the location information pertaining to the buggy code that requires to be fixed. 

Specifically, this paper approaches the bug-fixing problem as a conversation-driven text-generation task leveraging the advancements in LLMs. Unlike existing LLM-based bug-fixing techniques, which generate candidate patches directly from the given buggy code, \approach enhances the patch generation process with augmented buggy content (as shown in the upper part of Figure~\ref{overview}), consisting of the \colorbox{light-red}{\textbf{buggy code snippet}} with additional \colorbox{light-yellow}{\textbf{dependence context}} (if exists) at the class and repository levels, the programmer's intent (i.e., \colorbox{light-green}{\textbf{commit message}}) conveyed in the corresponding GitHub commit, and \colorbox{light-blue}{\textbf{similar bug-fixing demonstrations}} retrieved from the collected historical code corpus. Regarding the extraction of the \colorbox{light-yellow}{\textbf{dependence context}}, we initially employ the static analysis tool Spoon \cite{pawlak2016spoon} to parse the associated code files (i.e., the buggy class file and other dependent class files pertinent to cross-file dependencies) into abstract syntax trees (ASTs). Subsequently, we extract the necessary contextual dependencies related to the buggy method via data-flow analysis (e.g., definition-use chains). While traversing the AST of the buggy class file, \approach gathers three types of information as class-level dependencies: the imports of project-specific library depended upon by the buggy method, the global variables (defined within the buggy class scope) utilized in the buggy method, and the signatures of methods invoked within the buggy method. For repository-level dependencies, we extract the global variables utilized within the buggy method, along with the signatures of methods invoked by the buggy method, from the dependent classes that are imported in the buggy class. Due to the input token length limitation of LLMs, we restrict the extraction to only one layer of cross-file dependencies. Formally, given a new single-hunk bug, we propose leveraging the LLMs within \approach to generate a candidate patch \(p\) for fixing the corresponding bug. The bug-fixing task is defined as \(p=\mathtt{LLM}(\mathbf{PROMPT})\). Let \(\mathbf{PROMPT}=\mathbb{I} \oplus \mathbb{C}\oplus \mathbb{F}\) be the input prompt, where \(\mathbb{I}\) denotes the system instruction for prompting \(\mathtt{LLM}\) to align the collaborative behavior of human programmers, \(\mathbb{C}\) denotes the augmented buggy content of the given single-hunk bug provided by \approach, \(\mathbb{F}\) denotes different dimensions of feedback information from earlier bug management stages, and \(\oplus\) denotes the concatenation operation. Given \(\mathbf{PROMPT}\), the goal of \(\mathtt{LLM}\) is to learn the conditional probability \(\mathcal{P}(p|\mathbf{PROMPT})\).

When engaged in a single-turn conversation, the system instruction and user-defined prompt are used as input to generate an assistant message with ChatGPT. The system instruction plays a crucial role in defining the behavior of the assistant, allowing the agents to simulate the corresponding programmer behaviors. The user-defined prompt serves as a means to convey requests or comments for the assistant to respond to. By utilizing specific prompts, \approach effectively aligns the collaborative abilities of the programmers. This enables an interactive bug-fixing process, where the outputs from earlier stages are used to construct the input prompts for subsequent stages. As a result, \approach optimizes the ability of LLMs to generate correct patches that fix the given bugs. While \approach is a general framework capable of being applied to various LLMs, this paper employs the state-of-the-art ChatGPT model, which is tailored for dialogue-based interactions.

\subsection{Bug Reporting}

During the initial phase of bug management, the tester identifies a bug within the source code and proceeds to file a detailed report elucidating the nature of the bug. In practice, bug reports play a crucial role in bug fixing as they provide the developer with essential information regarding the discovered bug. These specific details significantly assist the developer in resolving the bug \cite{zimmermann2010what,zou2020practitioners}. To simulate the tester's behavior, \approach is designed to generate an initial bug report that outlines the underlying cause of the given buggy code.

\begin{figure}[tbp]
  \centering
  \includegraphics[width=\textwidth]{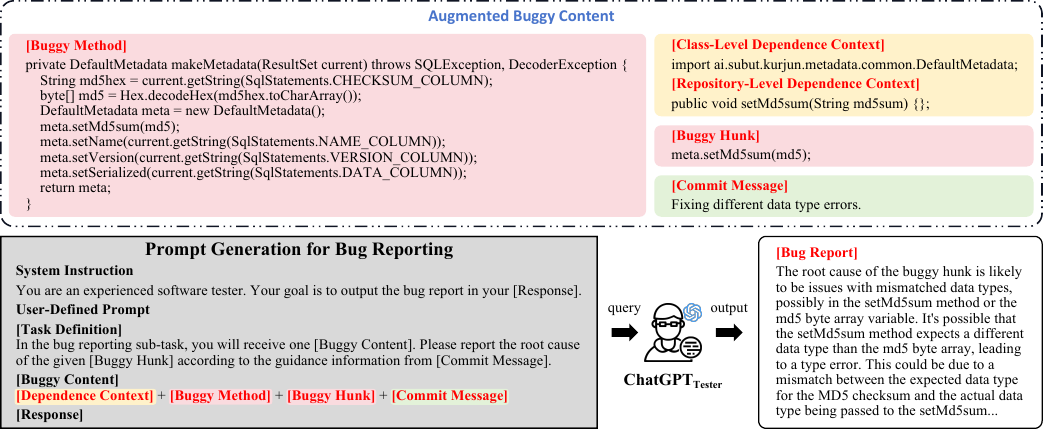}
  \caption{A Prompting Example of the Tester's Behavior during the Bug Reporting Stage.}
  \label{reporting}
  \Description{}
\end{figure}

Figure~\ref{reporting} illustrates the prompt generated to query $\mathbf{ChatGPT_{Tester}}$ during the bug reporting stage, along with the corresponding output. The \textbf{System Instruction} specifies the persona adopted by $\mathbf{ChatGPT_{Tester}}$ in its responses. The primary objective of $\mathbf{ChatGPT_{Tester}}$ is to report the root cause of the given \colorbox{light-red}{\color{red}{buggy method}} based on its fault location (i.e., \colorbox{light-red}{\color{red}{buggy hunk}}). To ensure a highly relevant response, the \textbf{User-Defined Prompt} provides crucial details and context to $\mathbf{ChatGPT_{Tester}}$, i.e., the definition of the bug reporting subtask and the augmented buggy content required for the bug reporting stage. Additionally, we use special delimiters (e.g., \textbf{[Task Definition]}) to explicitly indicate distinct parts of the \textbf{User-Defined Prompt}, facilitating $\mathbf{ChatGPT_{Tester}}$ to better comprehend the relationships between various contexts. Following the provided prompt, $\mathbf{ChatGPT_{Tester}}$ produces a \fcolorbox{black}{white}{\parbox{.09\linewidth}{\color{red}{bug report}}} that describes the bug's nature and its impact on the given \colorbox{light-red}{\color{red}{buggy method}} according to the guidance information from the \colorbox{light-yellow}{\color{red}{dependence context}} and the \colorbox{light-green}{\color{red}{commit message}}. As shown in the generated \fcolorbox{black}{white}{\parbox{.09\linewidth}{\color{red}{bug report}}}, $\mathbf{ChatGPT_{Tester}}$ correctly identifies the root cause of the \colorbox{light-red}{\color{red}{buggy hunk}} (i.e., \textit{It's possible that the setMd5sum method expects a different data type than the md5 byte array, leading to a type error.}). This information will then be assigned to the developer to assist in the resolution of the discovered bug during the subsequent stage.

\subsection{Bug Diagnosis}

Upon receiving a bug report from the tester, the developer commences the diagnosis process by utilizing the available information provided in the report. Practically, when assigned a newly discovered bug, the developer first engages in debugging practices. This involves carefully analyzing the source code line by line and documenting their findings in natural language. This self-guided approach enhances the efficiency of bug fixing without the need for external expert guidance \cite{chen2024teaching,paul2023automated}. Furthermore, the developer consults historical bug corpora to extract bug-fixing patterns that shed light on the causes and resolutions of similar issues. This mining process aids in acquiring valuable knowledge pertaining to the reasons behind bug occurrences and the corresponding fixes \cite{osman2014mining,tan2024crossfix}. To mimic the developer's diagnostic behavior, \approach initiates by employing rubber duck debugging techniques to provide a detailed, line-by-line explanation for the given buggy code snippet. Then, \approach retrieves relevant bug-fixing demonstrations for pattern summarization by analyzing the paired buggy and fixed methods. These two forms of guidance serve to aid the developer in generating the correct patches.

\begin{figure}[tbp]
  \centering
  \includegraphics[width=\textwidth]{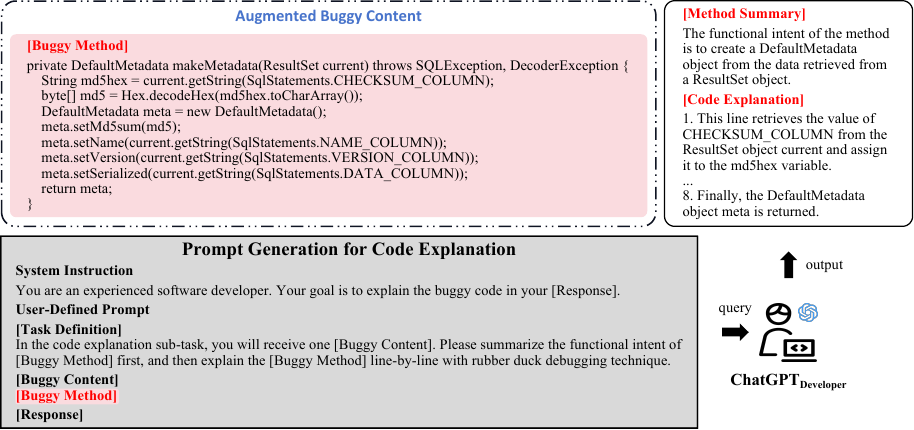}
  \caption{A Prompting Example of the Developer's Behavior during the Code Explanation Stage.}
  \label{explanation}
  \Description{}
\end{figure}

\begin{figure}[tbp]
  \centering
  \includegraphics[width=\textwidth]{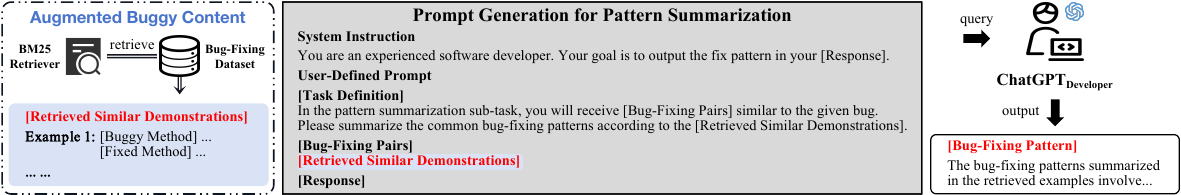}
  \caption{A Prompting Example of the Developer's Behavior during the Pattern Summarization Stage.}
  \label{summarization}
  \Description{}
\end{figure}

\subsubsection{Code Explanation}

Figure~\ref{explanation} illustrates an example prompt alongside the corresponding output obtained during the code explanation phase. The primary objective of $\mathbf{ChatGPT_{Developer}}$ is to elucidate the given \colorbox{light-red}{\color{red}{buggy method}} using the rubber duck debugging technique. This debugging process mimics a common practice employed by human programmers, which involves articulating their code line-by-line using natural language, as if conversing with a rubber duck \cite{spinellis2000pragmatic}. Inspired by the principle of rubber ducking, \approach specializes the persona of $\mathbf{ChatGPT_{Developer}}$ and its responses in the \textbf{System Instruction}. Unlike traditional line-by-line explanations, \approach prompts $\mathbf{ChatGPT_{Developer}}$ to explain the \colorbox{light-red}{\color{red}{buggy method}} with additional program context, i.e., functional description. According to the \textbf{[Task Definition]}, $\mathbf{ChatGPT_{Developer}}$ initially summarizes the intent of the \colorbox{light-red}{\color{red}{buggy method}}, providing a functional description to clarify the program's expected behavior. Subsequently, $\mathbf{ChatGPT_{Developer}}$ explains the code implementation of the \colorbox{light-red}{\color{red}{buggy method}} with the guidance of \fcolorbox{black}{white}{\parbox{.15\linewidth}{\color{red}{method summary}}} generated by $\mathbf{ChatGPT_{Developer}}$. This operation ensures that $\mathbf{ChatGPT_{Developer}}$ handles the \colorbox{light-red}{\color{red}{buggy method}} within a sufficiently concise context. Consequently, $\mathbf{ChatGPT_{Developer}}$ is able to provide a detailed \fcolorbox{black}{white}{\parbox{.145\linewidth}{\color{red}{code explanation}}} of the buggy content, thereby enhancing the efficiency of debugging without relying on additional artifacts such as unit tests.

\subsubsection{Pattern Summarization}
\label{patsum}

As depicted in Figure~\ref{summarization}, the main goal of $\mathbf{ChatGPT_{Developer}}$ at this phase is to summarize bug-fixing patterns by analyzing similar demonstrations retrieved from the historical code corpus. The initial step involves retrieving similar programs from a collected bug-fixing dataset based on the given buggy content. To achieve this, \approach employs the sparse keyword-based BM25 score \cite{robertson2009probabilistic} as the retrieval metric. The BM25 score, a probabilistic model widely used in previous studies \cite{wei2020retrieve,li2021editsum}, operates as a bag-of-words retriever, estimating the lexical-level similarity between two sentences. Higher BM25 scores indicate greater similarity between the sentences. Let \(\mathcal{D}=\{(B_i, F_i, L_i, C_i)\}_{i=1}^{|\mathcal{D}|}\) represent a bug-fixing dataset containing \(|\mathcal{D}|\) 4-tuple bug-fixing pairs, where \(B_i\) denotes the i-th buggy method, \(F_i\) denotes its paired fixed version, \(L_i\) denotes the location of the corresponding bug (i.e., the buggy hunk), and \(C_i\) denotes the mined GitHub commit message related to the bug. Specifically, \approach respectively retrieves the most relevant bug-fixing pair from \(\mathcal{D}\) based on a multi-faceted buggy context \(\mathcal{C} \leftarrow \{b, l, c\}\), where \(b\) denotes the given buggy method, \(l\) denotes its buggy hunk, and \(c\) denotes the commit message. For instance, \approach selects the top-1 retrieved output from the BM25 retriever based on the computed similarity score \(f(B_i, b)\) when considering the context of buggy method, where \(f\) denotes the relevance scoring function. To ensure retrieval performance, we further establish a dynamic threshold criteria: each retrieved bug-fixing pair must have a BM25 score greater than the length of the input query (i.e., the given buggy method, buggy hunk, or commit message) with an added term corresponding to the average length of \(\mathcal{C}\) across all bug-fixing pairs in \(\mathcal{D}\). In other words, the top-1 output is retained only if \(f(B_i, b) > len(b) + \frac{\sum_{i=1}^{|\mathcal{D}|} len(B_i)}{|\mathcal{D}|}\). Specifically, \approach selects up to three demonstrations (including paired buggy and fixed methods) as the retrieved results from \(\mathcal{D}\), taking into account the potential for duplicates when evaluating different buggy contexts, or excluding outputs that do not meet the threshold criteria. $\mathbf{ChatGPT_{Developer}}$ is then able to summarize the common \fcolorbox{black}{white}{\parbox{.16\linewidth}{\color{red}{bug-fixing patterns}}} based on \colorbox{light-blue}{\color{red}{retrieved similar demonstrations}}, providing insights into the root causes and resolutions of the given bug.

\subsection{Patch Generation}

Once the root cause of a bug has been identified, the developer embarks on the process of creating a patch to resolve the discovered bug. Previous studies have employed LLMs to directly generate candidate patches based on the provided buggy code. However, bug fixing is an intricate task that poses significant challenges in generating the correct patches from scratch. In the context of \approach, the responsibilities of the developer encompass two main aspects. Firstly, the developer generates an initial candidate patch according to the bug report filed by $\mathbf{ChatGPT_{Tester}}$ and the guidance obtained during the bug diagnosis stage. Secondly, the developer refines the candidate patch by incorporating review feedback when the candidate patch does not meet the desired fixing goal. This subsection exemplifies the process of initial patch generation.

\begin{figure}[tbp]
  \centering
  \includegraphics[width=\textwidth]{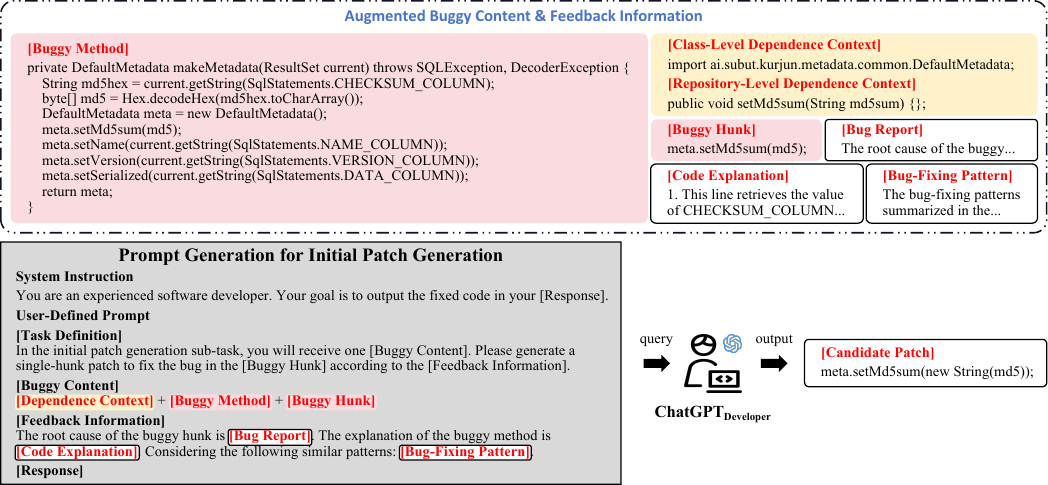}
  \caption{A Prompting Example of the Developer's Behavior during the Initial Patch Generation Stage.}
  \label{initial}
  \Description{}
\end{figure}

As illustrated in Figure~\ref{initial}, the objective of $\mathbf{ChatGPT_{Developer}}$, as stated in the \textbf{System Instruction}, during the initial patch generation stage is to patch the buggy hunk using the feedback information outlined in the \textbf{User-Defined Prompt}. Existing LLM-based approaches have encountered challenges regarding the accuracy of patch generation. In light of this, \approach provides $\mathbf{ChatGPT_{Developer}}$ with a guided process for bug reporting and diagnosis. This enables $\mathbf{ChatGPT_{Developer}}$ to generate an initial \fcolorbox{black}{white}{\parbox{.135\linewidth}{\color{red}{candidate patch}}} by incorporating information from the \fcolorbox{black}{white}{\parbox{.09\linewidth}{\color{red}{bug report}}}, the \fcolorbox{black}{white}{\parbox{.145\linewidth}{\color{red}{code explanation}}}, and the \fcolorbox{black}{white}{\parbox{.16\linewidth}{\color{red}{bug-fixing patterns}}} as prompt. Subsequently, the generated \fcolorbox{black}{white}{\parbox{.135\linewidth}{\color{red}{candidate patch}}} undergoes further verification by the reviewer.

\begin{algorithm}[tbp]
	\caption{Interaction Process between \(\mathbf{ChatGPT_{Developer}}\) and \(\mathbf{ChatGPT_{Reviewer}}\).}
	\label{algorithm1}
	\KwIn{\(\mathbf{ChatGPT_{Developer}}\) (the developer ChatGPT agent);\(\mathbf{ChatGPT_{Reviewer}}\) (the reviewer ChatGPT agent); \(p_{\mathbf{init}}\) (the candidate patch generated by \(\mathbf{ChatGPT_{Developer}}\) during the initial patch generation stage); \(\mathbf{PROMPT_{\mathbf{PV}}}\) (the patch verification prompt); \(\mathbf{PROMPT_{\mathbf{PG_{iter}}}}\) (the iteration patch generation prompt); \(\mathbf{maxIterNum}\) (the maximum iteration number)}
	\KwOut{\(p\) (the candidate patch verified by \(\mathbf{ChatGPT_{Reviewer}}\))}  
	\BlankLine
	\(\mathbf{currentIterNum} \leftarrow 0\)
	
	\(\mathbf{C_{review}} \leftarrow \mathbf{ChatGPT_{Reviewer}}(\mathbf{PROMPT_{\mathbf{PV}}}(p_{\mathbf{init}}))\)
	
	\uIf{\(\mathbf{C_{review}}\) is \(\mathtt{PASS}\)}{\(p \leftarrow p_{\mathbf{init}}\)}
	\Else{
	\While{\(\mathbf{currentIterNum} < \mathbf{maxIterNum}\)}{
	\(\mathbf{currentIterNum} \leftarrow \mathbf{currentIterNum} + 1\)\\
	\(p_{\mathbf{iter}} \leftarrow \mathbf{ChatGPT_{Developer}}(\mathbf{PROMPT_{\mathbf{PG_{iter}}}}(\mathbf{C_{review}}))\)\\
	\(\mathbf{C_{review}} \leftarrow \mathbf{ChatGPT_{Reviewer}}(\mathbf{PROMPT_{\mathbf{PV}}}(p_{\mathbf{iter}}))\)\\
	\uIf{\(\mathbf{C_{review}}\) is \(\mathtt{PASS}\)}{\(p \leftarrow p_{\mathbf{iter}}\)\\\(\mathbf{break}\)}
	}
	}
	\algorithmicreturn{ \(p\)}
\end{algorithm}

\subsection{Patch Verification}

Generating the correct patches in a single attempt presents a significant challenge for complex bugs. Therefore, the review process for the generated candidate patches assumes considerable significance as a pivotal activity within software peer review \cite{rigby2012contemporary,wang2015comparative}. When presented with a candidate patch generated by the developer, the reviewer needs to assess its effectiveness in resolving the given bug. Algorithm~\ref{algorithm1} details the interaction process between the developer and the reviewer. The inputs include two ChatGPT agents (i.e., $\mathbf{ChatGPT_{Developer}}$ and $\mathbf{ChatGPT_{Reviewer}}$), the initial candidate patch \(p_{\mathbf{init}}\) generated by \(\mathbf{ChatGPT_{Developer}}\), the patch verification prompt \(\mathbf{PROMPT_{\mathbf{PV}}}\), the iteration patch generation prompt \(\mathbf{PROMPT_{\mathbf{PG_{iter}}}}\), and the hyper-parameter for the maximum iteration number \(\mathbf{maxIterNum}\). The algorithm produces as output the final candidate patch \(p\), which has been verified by $\mathbf{ChatGPT_{Reviewer}}$. The algorithm begins by initializing the current iteration turn \(\mathbf{currentIterNum}\) (Line 1). Subsequently, $\mathbf{ChatGPT_{Reviewer}}$ generates the review feedback \(\mathbf{C_{review}}\) for \(p_{\mathbf{init}}\) using \(\mathbf{PROMPT_{\mathbf{PV}}}\) (Line 2). If $\mathbf{ChatGPT_{Reviewer}}$ passes \(p_{\mathbf{init}}\) (Line 3), the algorithm considers the bug resolved, and \(p_{\mathbf{init}}\) is regarded as the final output (Line 4). In cases where $\mathbf{ChatGPT_{Reviewer}}$ does not approve \(p_{\mathbf{init}}\), indicating the bug remains unfixed, an interactive process is initiated between $\mathbf{ChatGPT_{Reviewer}}$ and $\mathbf{ChatGPT_{Developer}}$. The hyper-parameter \(\mathbf{maxIterNum}\) serves as a termination criterion, capping the maximum number of iterations allowed to generate a candidate patch for fixing the bug (Line 6). During each iteration, \(\mathbf{currentIterNum}\) is incremented (Line 7). To generate an updated iteration candidate patch \(p_{\mathbf{iter}}\), $\mathbf{ChatGPT_{Developer}}$ integrates \(\mathbf{PROMPT_{\mathbf{PG_{iter}}}}\) with \(\mathbf{C_{review}}\) together as input (Line 8). $\mathbf{ChatGPT_{Reviewer}}$ then interactively evaluates \(p_{\mathbf{iter}}\) to assess its suitability as a correct solution to the bug. This interactive process continues until $\mathbf{ChatGPT_{Reviewer}}$ determines \(p_{\mathbf{iter}}\) to be correct (Lines 10--12) or when the maximum number of iterations \(\mathbf{maxIterNum}\) is reached.

\begin{figure}[tbp]
  \centering
  \includegraphics[width=\textwidth]{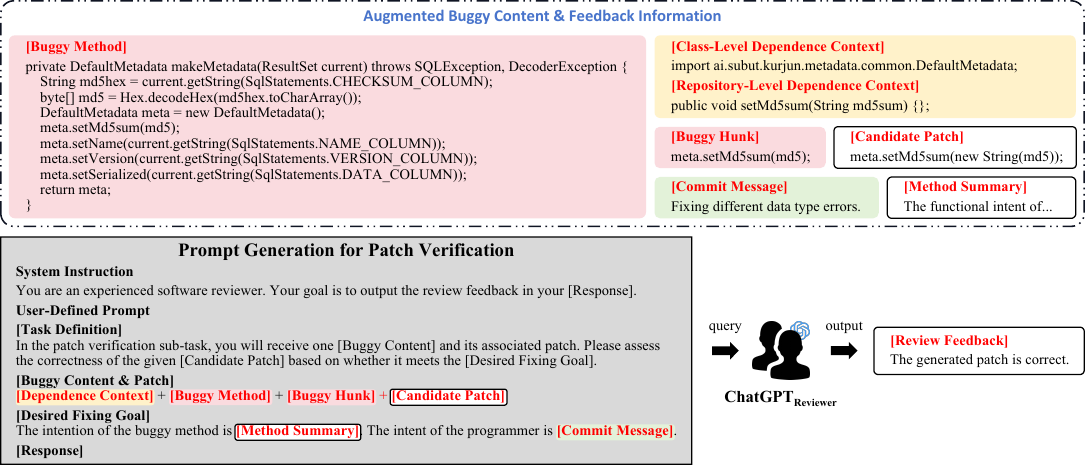}
  \caption{A Prompting Example of the Reviewer's Behavior during the Patch Verification Stage.}
  \label{verification}
  \Description{}
\end{figure}

Figure~\ref{verification} delineates the review process undertaken by $\mathbf{ChatGPT_{Reviewer}}$. The objective of $\mathbf{ChatGPT_{Reviewer}}$ during the patch verification stage is to evaluate the given \textbf{[Buggy Content \& Patch]}, determining its correctness based on the provided \textbf{[Desired Fixing Goal]}. The fixing goal encompasses two primary aspects: the functionality requirement (i.e., the \fcolorbox{black}{white}{\parbox{.15\linewidth}{\color{red}{method summary}}} inferred by $\mathbf{ChatGPT_{Developer}}$) and the programmer's intent as conveyed in the \colorbox{light-green}{\color{red}{commit message}}. As depicted in Figure~\ref{verification}, the \fcolorbox{black}{white}{\parbox{.135\linewidth}{\color{red}{review feedback}}} confirms that the \fcolorbox{black}{white}{\parbox{.135\linewidth}{\color{red}{candidate patch}}} satisfies both of the aforementioned goals, thereby deeming it correct.

\section{Experimental Setup}
\label{exp}

\subsection{Research Questions}

To assess the effectiveness of \approach, we aim at answering the following three research questions (RQs): 
\begin{itemize}
  \item \textbf{RQ1: How does \approach perform in bug fixing when compared to state-of-the-art LLMs?} The objective of this RQ is to evaluate the superior effectiveness of \approach in comparison to state-of-the-art LLM baselines within the context of bug fixing. To achieve this, we conduct a comprehensive comparison of \approach against 13 LLMs using an augmented bug-fixing benchmark.
  \item \textbf{RQ2: How does each component impact the performance of \approach?} \approach introduces two essential mechanisms: the augmented buggy content and the behavior-simulation framework. The proposed framework further comprises three agents: $\mathbf{ChatGPT_{Tester}}$ is responsible for bug reporting, $\mathbf{ChatGPT_{Developer}}$ handles bug diagnosis and patch generation, and $\mathbf{ChatGPT_{Reviewer}}$ is in charge of patch verification. In this RQ, we aim to analyze the contributions of each designed component by conducting the ablation study.
  \item \textbf{RQ3: What is the generalizability of \approach to additional benchmarks and different LLMs?} This RQ first assesses the generalizability of \approach by applying it to four common benchmarks in the field of automated program repair (APR). Furthermore, we extend \approach to five open-source LLMs that support interactive dialogues to enhance evaluation diversity.
\end{itemize}

\subsection{Benchmark}

In this paper, we utilize the widely recognized BFP benchmark \cite{tufano2019empirical} as our original data source, encompassing a substantial collection of paired bug-fixing instances extracted from real-world GitHub repositories. Each instance within the BFP benchmark consists of both the buggy and the fixed Java methods. To achieve a comprehensive understanding of the provided buggy method, we first extract additional contextual dependencies associated with the buggy method as described in Section~\ref{tafo}. Notably, we exclude instances that cannot be successfully parsed by Spoon, as well as those exceeding 300 tokens in length (in consideration of the maximum token limits of the LLMs). Furthermore, we collect bug-fixing commits from GitHub linked to instances within the BFP benchmark and apply a three-stage filtering mechanism to ensure commit quality. First, we filter out commits shorter than 5 tokens, excluding low-informative messages such as ``done'' or ``bug fixing''. The remaining commits undergo an automated annotation process powered by the advanced LLM GPT-4 \cite{he2024crowdsourced}. Specifically, we provide GPT-4 with examples of irrelevant or low-quality commits to facilitate few-shot in-context learning, enabling it to accurately label the commits as either \textit{good} or \textit{bad}, and to generate a confidence score for each annotation. Finally, we employ random sampling \cite{singh2014sampling} on the \textit{good} commits with a confidence score greater than 0.90 for manual inspection. In particular, the first author and three experienced master students manually reviewed 380 commit samples (confidence level: 95\%, margin of error: 5\%). The inspection results show that GPT-4’s annotation accuracy exceeds 90\%, with 36 commits removed due to the inconsistencies between human and GPT-4 annotations, thereby demonstrating the effectiveness and reliability of the data annotations generated by GPT-4.

\begin{table}[tbp]
    \centering
    \caption{Statistics of the BFP Benchmark in Our Experiments.}
    \label{statistics}
    \resizebox{\textwidth}{!}{
    \begin{tabular}{rrccrcccr}
        \toprule
        \multirow{2.5}{*}{\textbf{}} & \multirow{2.5}{*}{\textbf{\# of Instances}} & \multicolumn{3}{c}{\textbf{Buggy Method Statistics}} & \multicolumn{4}{c}{\textbf{Buggy Class Statistics}} \\
        \cmidrule[0.5pt](rl){3-5}\cmidrule[0.5pt](rl){6-9}
        & & \textbf{LoC Range} & \textbf{CC Range} & \textbf{Complexity} & \textbf{LoC Range} & \textbf{Method Count Range} & \textbf{CC Range} & \textbf{Complexity} \\
        \midrule
        Training Set & 28226 & [1, 85] & [1, 56] & 17.2\% & [5, 63808] & [1, 2218] & [1, 49] & 11.7\% \\
       Testing Set & 3112 & [1, 57] & [1, 19] & 14.4\% & [5, 13765] & [1, 449] & [1, 27] & 9.8\% \\
        \bottomrule
    \end{tabular}}
\end{table}

Consequently, each instance within the BFP benchmark is augmented with additional information (i.e., the dependence context and the commit message). Next, we split the augmented BFP benchmark into training and testing sets by maintaining a 9:1 ratio. To prevent any data leakage, instances originating from the same GitHub repository are not allowed to appear in different sets (e.g., one in the training set and the other in the testing set). As illustrated in the second column of Table~\ref{statistics}, we collect 28226 bug-fixing instances in the training set and 3112 in the testing set. The \textbf{LoC Range} column specifies the range of lines of code (LoC) for both the buggy method and its corresponding buggy class. The \textbf{CC Range} column represents the range of cyclomatic complexity (CC) \cite{mcCabe1976complexity}, while the \textbf{Complexity} column indicates the proportion of complex methods with a CC greater than 5 \cite{yamashita2016thresholds}. The \textbf{Method Count Range} shows the range of method counts within the corresponding buggy class file. We evaluate the bug-fixing performance of \approach and selected baselines on the testing set. Additionally, we perform the pattern summarization process, as described in Section~\ref{patsum}, by retrieving similar instances from the training set. In real-world development scenarios, such a code corpus for retrieval can be constructed using open-source code repositories or historical bug-fixing data collected from developers' private projects to facilitate the pattern summarization process.

\subsection{Baselines}

This paper centers on addressing the bug-fixing task using LLMs. Therefore, we compare \approach against 13 state-of-the-art LLM baselines as listed in Table~\ref{benchmark}. The selection criteria are as follows:
\begin{itemize}
  \item \textbf{Popularity.} Initially, we consider the list of popular models hosted on the Hugging Face platform, which is an open-source resource for hosting and deploying large models. From this repository, we select LLMs pre-trained or fine-tuned on a large amount of code corpus and specifically engineered to address code-related tasks. Furthermore, we include closed-source LLMs (i.e., the GPT family of models) due to their demonstrated impressive performance across a wide range of tasks.
  \item \textbf{Diversity.} We select LLMs with varying sizes of parameters and from different organizations.
  \item \textbf{Accessibility.} The selected LLMs are publicly accessible either through checkpoints (e.g., InCoder) or non-free APIs (e.g., GPT-4). Thus, closed-source models such as AlphaCode \cite{li2022competition} are excluded from our evaluation.
\end{itemize}

\begin{table}[ht]
    \centering
    \caption{Overview of the Selected LLM Baselines.}
    \label{benchmark}
    \begin{tabular}{lrcc}
        \toprule
        \multicolumn{1}{c}{\textbf{Model}} & \multicolumn{1}{c}{\textbf{\# of Parameters}} & \textbf{Organization} & \textbf{Pre-Training / Fine-Tuning Code Corpus} \\
        \midrule
        CodeGPT \cite{lu2021codexglue} & 124M & Microsoft & CodeSearchNet \cite{husain2019codesearchnet} \\
        DeepSeek-Coder \cite{guo2024deepseek} & 1.3B & DeepSeek & GitHub \\
        CodeGen2 \cite{nijkamp2023codegen2} & 3.7B & Salesforce & Stack \cite{kocetkov2023stack} \\
        CodeGeeX2 \cite{zheng2023codegeex} & 6B & THU & Pile \cite{gao2021pile} \& BigQuery \& GitHub \\
        InCoder \cite{fried2023incoder} & 6.7B & Facebook & StackOverFlow \& GitHub \& GitLab \\
        Mistral \cite{jiang2023mistral} & 7B & Mistral\_AI & Hugging Face Repository\\
        CodeLLaMA \cite{roziere2023code} & 7B \& 13B & Meta & BigQuery \\
        StarCoder \cite{li2023starcoder} & 15B & BigCode & Stack \cite{kocetkov2023stack} \\
        GPT-NeoX \cite{black2022gptneox} & 20B & EleutherAI & Pile \cite{gao2021pile} \\
        Codex \cite{chen2021evaluating} & 175B & OpenAI & - \\
        ChatGPT \cite{openai2022chatgpt} & - & OpenAI & - \\
        GPT-4 \cite{openai2023gpt4} & - & OpenAI & - \\
        \bottomrule
    \end{tabular}
\end{table}

\subsection{Metrics}

This paper employs the following two evaluation metrics to compare the performance of \approach with the LLM baselines.
\begin{itemize}
  \item \textbf{Fix@k.} This paper first utilizes the Fix@k metric \cite{zhong2022standupnpr} to evaluate the bug-fixing performance of LLMs on the testing set. This metric measures the percentage of successfully resolved bugs within the entire testing set when \(k\) candidate patches are generated for each given bug. In other words, given a Java method with a single-hunk bug, each corresponding LLM is permitted to generate $k$ candidate patches. The bug is considered resolved if any of the LLM-generated patches exactly match the human-written ground truth. For Fix@k, higher values indicate better performance. In our experiments, we evaluate Fix@k with $k$ set to 1, 3, and 5, since most programmers are usually willing to review as few patches as possible in real-world development scenarios \cite{noller2022trust}.
  \item \textbf{Levenshtein Distance.} The Levenshtein distance metric calculates the absolute token-based edit distance between the LLM-generated candidate patch and the human-written ground truth, representing the minimum number of operations required to transform the candidate patch into the ground truth. A lower Levenshtein distance signifies a closer correspondence between the candidate patch and the ground truth. This metric serves as a complementary measure that provides valuable insights into the usefulness of incorrect LLM-generated candidate patches for programmers.
\end{itemize}

\subsection{Implementation}
\label{implementation}

We implement the main logic of \approach in Python by invoking ChatGPT through its API. Specifically, we employ the \textbf{gpt-3.5-turbo-0125} version of the ChatGPT family due to its enhanced performance and cost-efficiency. Following the best-practice guidelines from Shieh et al. \cite{shieh2023best}, we design prompts and manually examine several alternative approaches with selected buggy code using the Web-version of ChatGPT. The response configurations of each ChatGPT agent are detailed as follows:
\begin{itemize}
  \item $\mathbf{ChatGPT_{Tester}.}$ The maximum length of the \fcolorbox{black}{white}{\parbox{.095\linewidth}{\color{red}{bug report}}} generated during the bug reporting stage is restricted to 200 tokens.
  \item $\mathbf{ChatGPT_{Developer}.}$ The maximum length for the \fcolorbox{black}{white}{\parbox{.155\linewidth}{\color{red}{code explanation}}} and the \fcolorbox{black}{white}{\parbox{.165\linewidth}{\color{red}{bug-fixing pattern}}} generated during the bug diagnosis stage is limited to 500 tokens. During the patch generation stage, the maximum length of the \fcolorbox{black}{white}{\parbox{.14\linewidth}{\color{red}{candidate patch}}} is capped at 150 tokens.
  \item $\mathbf{ChatGPT_{Reviewer}.}$ The maximum length of the \fcolorbox{black}{white}{\parbox{.145\linewidth}{\color{red}{review feedback}}} generated during the patch verification stage is constrained to 200 tokens.
\end{itemize}
In our experiments, when $k=1$, we utilize greedy decoding for each ChatGPT agent. Specifically, \approach produces the top-1 chat completion choice for each input query with a sampling temperature of 0. For $k>1$, a sampling temperature of 0.8 is used to generate multiple responses. We limit the maximum number of interaction turns between $\mathbf{ChatGPT_{Reviewer}}$ and $\mathbf{ChatGPT_{Developer}}$ to three, as recommended by Chen et al. \cite{chen2024teaching}. Furthermore, we conduct experiments under a zero-shot setting, where task examples are not provided, aiming to demonstrate the superiority of \approach.

\section{Results and Analysis}
\label{res}

\begin{figure}[t]
  \centering
  \includegraphics[width=0.87\textwidth]{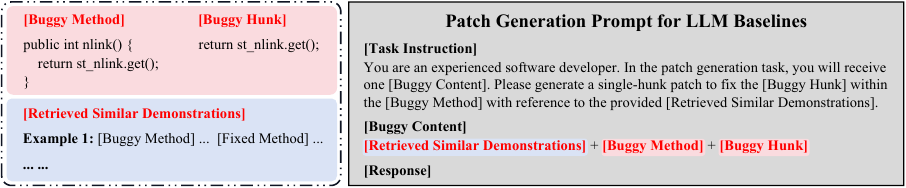}
  \caption{A Prompting Example for the LLM Baselines.}
  \label{base-prompt}
  \Description{}
\end{figure}

\subsection{Answering RQ1}

To answer this question, we conduct a comprehensive comparison of \approach with 13 state-of-the-art baselines on the augmented BFP benchmark. For each baseline, we either reuse the official checkpoint or access the inference API for implementation. As illustrated in Figure~\ref{base-prompt}, we prompt the LLM baselines by incorporating both the buggy method and several task examples, following the approach established in prior research. To ensure a fair comparison, we utilize the same bug-fixing demonstrations retrieved by \approach during the pattern summarization phase to conduct these few-shot setting experiments. Additionally, we employ the same sampling hyper-parameters as \approach.

\begin{table}[t]
    \centering
    \caption{Comparison of \approach against the LLM Baselines.}
    \label{rq1a}
    \resizebox{\textwidth}{!}{
    \begin{tabular}{lrrrrrrr}
        \toprule
        \multirow{2.5}{*}{\textbf{Model}} & \multirow{2.5}{*}{\textbf{\# of Parameters}} & \multicolumn{4}{c}{\textbf{Evaluation Metric Result}} & \multicolumn{2}{c}{\textbf{Paired t-test Result}} \\
        \cmidrule[0.5pt](rl){3-6}\cmidrule[0.5pt](rl){7-8}
        & & \textbf{Fix@1 (\%) $\uparrow$} & \textbf{Fix@3 (\%) $\uparrow$} & \textbf{Fix@5 (\%) $\uparrow$} & \textbf{Levenshtein Distance $\downarrow$} & \textbf{T-statistic} & \textbf{$p$-value} \\
        \midrule
        CodeGPT & 124M & 2.76 & 5.46 & 6.65 & 73.69 & 53.90 & < 0.001 \\
        DeepSeek-Coder & 1.3B & 8.42 & 11.25 & 14.14 & 48.34 & 316.67 & < 0.001 \\
        CodeGen2 & 3.7B & 4.31 & 7.23 & 10.15 & 70.13 & 469.32 & < 0.001 \\
        CodeGeeX2 & 6B & 10.93 & 15.30 & 19.41 & 40.18 & 28.52 & < 0.001 \\
        InCoder & 6.7B & 5.24 & 8.03 & 11.60 & 64.20 & 117.10 & < 0.001 \\
        Mistral & 7B & 9.38 & 14.11 & 16.68 & 43.45 & 45.72 & < 0.001 \\
        CodeLLaMA & 7B & 9.45 & 14.46 & 17.96 & 43.35 & 28.98 & < 0.001 \\
        CodeLLaMA & 13B & 9.80 & 14.62 & 17.45 & 41.20 & 39.15 & < 0.001 \\
        StarCoder & 15B & 13.50 & 17.58 & 20.89 & 34.27 & 43.42 & < 0.001 \\
        GPT-NeoX & 20B & 8.93 & 14.56 & 16.16 & 57.37 & 32.57 & < 0.001 \\
        Codex & 175B & 11.05 & - & - & 36.33 & - & - \\
        ChatGPT & - & 14.62 & 18.99 & 22.27 & 33.28 & 43.21 & < 0.001 \\
        GPT-4 & - & 19.96 & 24.42 & 26.00 & 26.07 & 32.08 & < 0.001 \\
        \midrule
        \approach & - & \textbf{33.97 {\color{red}(14.01 $\uparrow$)}} & \textbf{37.08 {\color{red}(12.66 $\uparrow$)}} & \textbf{39.81 {\color{red}(13.81 $\uparrow$)}} & \textbf{21.44  {\color{red}(4.63 $\downarrow$)}} \\
        \bottomrule
    \end{tabular}}
\end{table}

\subsubsection{Experimental Metric Evaluation}

Table~\ref{rq1a} presents the bug-fixing performance of different models in terms of the \textbf{Fix@k} ($k \in [1,3,5]$) and \textbf{Levenshtein Distance} (when $k=1$) metrics, with the best result for each metric highlighted in bold. The numbers in {\color{red}red} denote \approach's improvement percentage points compared to the best baseline. Our experiments reveal the following three-fold key findings:
\begin{enumerate}
	\item \textbf{\approach demonstrates superior performance compared to all the baselines on the augmented BFP benchmark.} Specifically, \approach outperforms the best baseline GPT-4 by 14.01 percentage points in terms of \textbf{Fix@1}. These improvements underscore the effectiveness of \approach in bug fixing, particularly considering that \textbf{Fix@1} is a stringent metric. As for the \textbf{Levenshtein Distance} metric, \approach achieves a score of 21.44 on the augmented BFP benchmark, showcasing an improvement of 4.63 percentage points over GPT-4. We also perform a statistical comparison of performance in terms of \textbf{Fix@k} ($k \in [1,3,5]$) between \approach and each baseline LLM using the paired t-test \cite{dyba2006systematic}. The paired t-test evaluates the null hypothesis, which posits that the difference between \approach and each baseline is not statistically significant. If the reported $p$-value is less than the significance level of 0.05, the null hypothesis is rejected, indicating that the observed disparity between \approach and each baseline is statistically significant and not due to random chance. Additionally, we compute the T-statistic to measure the effect size; a larger T-statistic indicates a more significant performance difference between \approach and each baseline. The paired t-test results in Table~\ref{rq1a} show that \approach surpasses all baselines in bug-fixing performance, with the difference being statistically significant ($p$-value < 0.001).
	\item \textbf{Simulating programmer behavior proves to be advantageous for bug fixing.} Rather than altering the parameters of ChatGPT, \approach explicitly prompts ChatGPT to mimic the behavior of programmers engaged in the bug management process. The substantial improvements observed over the base ChatGPT model suggest that \approach effectively enhances ChatGPT's bug-fixing capabilities by endowing it with collaborative problem-solving skills. Furthermore, aligning ChatGPT with the interactive decision-making processes of programmers boosts its performance across various aspects, including the comprehension of intent.
	\item \textbf{Enhancing the bug-fixing performance of LLMs relies on two factors: increasing the number of parameters and designing well-crafted prompts.} Generally, an increase in parameters often leads to improved performance, as demonstrated by StarCoder-15B surpassing CodeGeeX2-6B, while CodeLLaMA-13B performs better than CodeLLaMA-7B. Notably, LLMs struggle to achieve satisfactory performance due to their inability to comprehend how to solve complex problems. However, this limitation can be effectively addressed by incorporating essential information into the query prompts. By adopting this approach, \approach significantly outperforms the LLM baselines. This finding validates our motivation to decompose the bug-fixing task into subtasks using well-designed prompts, as it substantially enhances the performance of LLMs in this context.
\end{enumerate}

\subsubsection{Bug Types Evaluation}

We begin by utilizing the Gumtree algorithm \cite{falleri2014fine} to compute the difference between two ASTs generated by Spoon \cite{pawlak2016spoon}. We classify the bugs into four categories based on the necessary edit operations to transform a bug hunk into its corrected version: \textbf{Simple Delete}, \textbf{Simple Insert}, \textbf{Simple Replace}, and \textbf{Mixed}. For instance, \textbf{Simple Delete} indicates that a bug can be fixed by removing certain tokens from a specific position. Table~\ref{rq1b} presents the comparison results (when $k=1$ and $k=5$), with each row representing a model and the corresponding number of correct patches generated for each bug type on the augmented BFP benchmark. The best results are highlighted in bold. The statistical findings in Table~\ref{rq1b} reveal that LLMs excel at fixing bugs that require only deletion edit operations but encounter difficulties with more intricate ones. This observation is reasonable since insert and replace edit operations demand that the model search for additional tokens to fix the given bug. Overall, \approach outperforms all baselines in resolving both simple and complex bugs.

\begin{table}[tbp]
\centering
    \caption{Comparison of the Fixed Bug Types between \approach and the LLM Baselines.}
    \label{rq1b}
    \resizebox{\textwidth}{!}{
    \begin{tabular}{lrrrrrrrr}
    \toprule
    \multirow{3.5}{*}{\textbf{Model}} & \multicolumn{2}{c}{\textbf{Simple Delete}} & \multicolumn{2}{c}{\textbf{Simple Insert}} & \multicolumn{2}{c}{\textbf{Simple Replace}} & \multicolumn{2}{c}{\textbf{Mixed}} \\
    \cmidrule[0.5pt](rl){2-3}\cmidrule[0.5pt](rl){4-5}\cmidrule[0.5pt](rl){6-7}\cmidrule[0.5pt](rl){8-9}
    & \multicolumn{2}{c}{149 bugs} & \multicolumn{2}{c}{272 bugs} & \multicolumn{2}{c}{1157 bugs} & \multicolumn{2}{c}{1534 bugs} \\
    & \multicolumn{1}{c}{$k=1$} & \multicolumn{1}{c}{$k=5$} & \multicolumn{1}{c}{$k=1$} & \multicolumn{1}{c}{$k=5$} & \multicolumn{1}{c}{$k=1$} & \multicolumn{1}{c}{$k=5$} & \multicolumn{1}{c}{$k=1$} & \multicolumn{1}{c}{$k=5$} \\
    \midrule
    CodeGPT & 15 (8.66\%) & 22 (14.77\%) & 5 (1.84\%) & 28 (10.29\%) & 36 (3.11\%) & 90 (7.78\%) & 30 (1.96\%) & 67 (4.37\%) \\
    DeepSeek-Coder & 28 (18.79\%) & 42 (28.19\%) & 14 (5.15\%) & 51 (18.75\%) & 112 (9.68\%) & 156 (13.48\%) & 108 (7.04\%) & 191 (12.45\%) \\
    CodeGen2 & 21 (14.09\%) & 38 (25.50\%) & 9 (3.31\%) & 38 (13.97\%) & 46 (3.98\%) & 100 (8.64\%) & 58 (3.78\%) & 140 (9.13\%) \\
    CodeGeeX2 & 23 (15.44\%) & 50 (33.56\%) & 51 (18.75\%) & 61 (22.43\%) & 129 (11.15\%) & 226 (19.53\%) & 137 (8.93\%) & 267 (17.41\%) \\
    InCoder & 21 (14.09\%) & 44 (29.53\%) & 14 (5.15\%) & 62 (22.79\%) & 59 (5.10\%) & 97 (8.38\%) & 69 (4.50\%) & 158 (10.30\%) \\
    Mistral & 29 (19.46\%) & 43 (28.86\%) & 16 (5.88\%) & 33 (12.13\%) & 124 (10.72\%) & 222 (19.19\%) & 123 (8.02\%) & 221 (14.41\%) \\
    CodeLLaMA-7B & 29 (19.46\%) & 49 (32.89\%) & 13 (4.78\%) & 26 (9.56\%) & 136 (11.75\%) & 239 (20.66\%) & 116 (7.56\%) & 245 (15.97\%) \\
    CodeLLaMA-13B & 28 (18.79\%) & 46 (30.87\%) & 14 (5.15\%) & 26 (9.56\%) & 137 (11.84\%) & 229 (19.79\%) & 126 (8.21\%) & 242 (15.78\%) \\
    StarCoder & 27 (18.12\%) & 55 (36.91\%) & 52 (19.12\%) & 62 (22.79\%) & 164 (14.17\%) & 252 (21.78\%) & 177 (11.54\%) & 281 (18.32\%) \\
    GPT-NeoX & 28 (18.79\%) & 59 (39.60\%) & 12 (4.41\%) & 59 (21.69\%) & 127 (10.98\%) & 146 (12.62\%) & 111 (7.24\%) & 239 (15.58\%) \\
    Codex & 19 (12.75\%) & - & 6 (2.21\%) & - & 175 (15.13\%) & - & 144 (9.39\%) & - \\
    ChatGPT & 27 (18.12\%) & 49 (32.89\%) & 15 (5.51\%) & 59 (21.69\%) & 218 (18.84\%) & 290 (25.06\%) & 195 (12.71\%) & 295 (19.23\%) \\
    GPT-4 & 38 (25.50\%) & 48 (32.21\%) & 38 (13.97\%) & 42 (15.44\%) & 277 (23.94\%) & 370 (31.98\%) & 268 (17.47\%) & 349 (22.75\%) \\
    \midrule
    \approach & \textbf{60 (40.27\%)} & \textbf{77 (51.68\%)} & \textbf{96 (35.29\%)} & \textbf{115 (42.28\%)} & \textbf{448 (38.72\%)} & \textbf{497 (42.96\%)} & \textbf{453 (29.53\%)} & \textbf{550 (35.85\%)} \\
    \bottomrule
    \end{tabular}}
\end{table}

\begin{figure}[tbp]
  \centering
  \includegraphics[width=\textwidth]{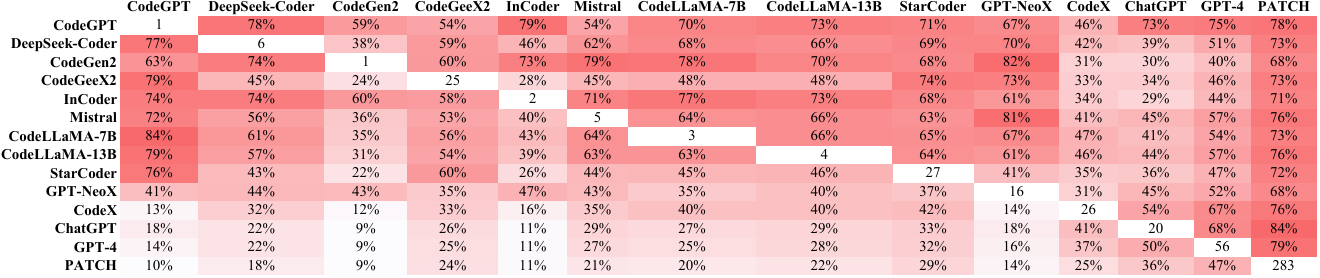}
  \caption{The Overlapping Rates and Unique Patch Numbers of the Evaluated Models.}
  \label{rq1c}
  \Description{}
\end{figure}

\subsubsection{Overlapping Phenomenon Evaluation}

As illustrated in Figure~\ref{rq1c}, each row represents the overlapping ratio (when $k=1$) of correct patches generated by one model with those generated by other models, while the diagonal indicates the number of unique correct patches generated by each model on BFP. The color intensity of each rectangle increases with the overlapping rate, providing a visual cue for easier interpretation. For example, \approach generates correct patches that overlap with 29\% of those generated by StarCoder (row 14, column 9). The results in Figure~\ref{rq1c} indicate that models with superior performance tend to exhibit higher overlapping patching rates with other models. The evaluation results in Table~\ref{rq1a} show that that \approach, GPT-4, and ChatGPT are the top three models. The overlapping rates of other models with these three are notably higher, likely due to the adoption of similar network architectures and inference paradigms among DL-based approaches. Furthermore, \approach uniquely fixes 283 bugs (row 14, column 14), a significantly higher number than other LLM baselines.

\begin{tcolorbox}[size=title]
    \textbf{Answer to RQ1}: In conclusion, \approach demonstrates significant superiority over the LLM baselines across the evaluation metrics, underscoring the effectiveness of \approach in the bug-fixing task. Additionally, our findings reveal that \approach is capable of generating a higher number of unique and correct patches compared to the LLM baselines.
\end{tcolorbox}

\subsection{Answering RQ2}

To answer this question, we perform a series of ablation experiments to assess the impact of different components within the \approach design. To ensure the fairness of comparisons, we maintained consistency in the implementation settings, adhering to the parameters detailed in Section~\ref{implementation}.

\begin{table}[tbp]
    \centering
    \caption{Ablation Study for \approach.}
    \label{rq2a}
    \resizebox{\textwidth}{!}{
    \begin{tabular}{cccccccrrr}
        \toprule
        \multirow{4}{*}{\textbf{Model}} & \multicolumn{6}{c}{\textbf{Component}} & \multirow{4}{*}{\textbf{Fix@1 (\%) $\uparrow$}} & \multirow{4}{*}{\textbf{Fix@3 (\%) $\uparrow$}} & \multirow{4}{*}{\textbf{Fix@5 (\%) $\uparrow$}} \\
        \cmidrule[0.5pt](rl){2-7}
        & \multicolumn{2}{c}{\textbf{Augmented Buggy Content}} & $\mathbf{ChatGPT_{Tester}}$ & \multicolumn{2}{c}{$\mathbf{ChatGPT_{Developer}}$} & $\mathbf{ChatGPT_{Reviewer}}$ \\
        \cmidrule[0.5pt](rl){2-3}\cmidrule[0.5pt](rl){4-4}\cmidrule[0.5pt](rl){5-6}\cmidrule[0.5pt](rl){7-7}
        & \texttt{Dependence Context} & \texttt{Commit Message} & \texttt{Bug Report} & \texttt{Code Explanation} & \texttt{Bug-Fixing Pattern} & \texttt{Review Feedback} \\
        \midrule
        \multirow{11}{*}{\textbf{ChatGPT}} & \faTimes & \faTimes & & & & & 14.62 & 18.99 & 22.27 \\
        & \faCheck & \faTimes & & & & & 16.87 & 21.50 & 24.39 \\
        & \faTimes & \faCheck & & & & & 18.86 & 24.33 & 27.28 \\
        & & & \faCheck & \faTimes & \faTimes & \faTimes & 17.48 & 22.59 & 26.86 \\
        & & & \faTimes & \faCheck & \faTimes & \faTimes & 17.26 & 22.46 & 26.09 \\
        & & & \faTimes & \faTimes & \faCheck & \faTimes & 16.93 & 21.79 & 24.55 \\
        & & & \faTimes & \faCheck & \faCheck & \faTimes & 17.89 & 23.01 & 27.37 \\
        & & & \faCheck & \faCheck & \faTimes & \faTimes & 22.14 & 26.22 & 29.53 \\
        & & & \faCheck & \faTimes & \faCheck & \faTimes & 20.31 & 25.16 & 28.92 \\
        & & & \faCheck & \faCheck & \faCheck & \faTimes & 24.23 & 27.79 & 30.27 \\
        & & & \faCheck & \faCheck & \faCheck & \faCheck & \textbf{33.97} & \textbf{37.08} & \textbf{39.81} \\
        \midrule
        \multirow{3}{*}{\textbf{GPT-4}} & \faTimes & \faTimes & & & & & 19.96 & 24.42 & 26.00 \\
        & \faCheck & \faTimes & & & & & 22.85 & 27.47 & 30.01 \\
        & \faTimes & \faCheck & & & & & 25.96 & 29.37 & 31.62 \\
        \bottomrule
    \end{tabular}}
\end{table}

\subsubsection{Ablation Study}

\begin{figure}[tbp]
  \centering
  \includegraphics[width=0.46\textwidth]{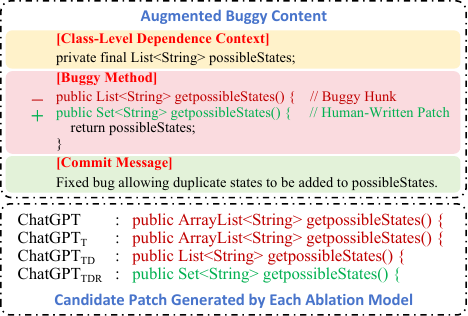}
  \caption{An Example from the Augmented BFP Benchmark only Fixed by \approach.}
  \label{rq2b}
  \Description{}
\end{figure}

Table~\ref{rq2a} presents the evaluation results, where each row represents one ablation model. The symbols \faCheck\ and \faTimes\ respectively indicate the addition and removal of the corresponding component. The best \textbf{Fix@k} ($k \in [1,3,5]$) results are highlighted in bold. To demonstrate the contribution of each component to bug-fixing performance, we start with evaluating the base model-ChatGPT-using only the buggy method as the input prompt to generate candidate patches. Upon integrating the \texttt{Dependence Context} and \texttt{Commit Message} into ChatGPT, we observe improvements of 2.25\% and 4.24\% in the \textbf{Fix@1} metric, respectively, compared to the base ChatGPT model. Further enhancements are observed when incorporating the \texttt{Bug Report}, \texttt{Code Explanation}, and \texttt{Bug-Fixing Pattern} components, which lead to additional improvements of 2.86\%, 2.64\%, and 2.31\% in \textbf{Fix@1}, respectively, relative to the base ChatGPT model. Notably, the addition of the $\mathbf{ChatGPT_{Tester}}$ component allows the ChatGPT agent to perform bug reporting, identify the root cause of the buggy code, and generate candidate patches based on the information available in the bug report. This improvement underscores the significance of providing essential bug-related information in the bug-fixing process. The inclusion of the $\mathbf{ChatGPT_{Developer}}$ component further enables the ChatGPT agent to generate candidate patches informed by the bug diagnosis stage, demonstrating the efficacy of the self-guided diagnosis process in enhancing bug-fixing efficiency. Furthermore, the integration of the $\mathbf{ChatGPT_{Reviewer}}$ component equips ChatGPT with interactive abilities to refine candidate patches based on \texttt{Review Feedback}, leading to continuous improvements in bug-fixing performance, with an additional enhancement of 9.74\% in \textbf{Fix@1}. This result emphasizes the importance of interaction and collaboration during the resolution of complex software bugs. In conclusion, all components are crucial for optimizing the bug-fixing performance of \approach. In the subsequent set of experiments, we further evaluate the impact of augmenting GPT-4, the most advanced LLM, with additional buggy content. Our results indicate that incorporating the \texttt{Dependence Context} and \texttt{Commit Message} components enhances GPT-4's bug-fixing performance by 2.89\% and 6.00\%, respectively, in terms of the \textbf{Fix@1} metric. This outcome indicates the significant positive effect of incorporating \texttt{Augmented Buggy Content}, which is on par with the performance observed in ChatGPT when utilizing the $\mathbf{ChatGPT_{Tester}}$ and $\mathbf{ChatGPT_{Developer}}$ components. Additionally, this result also highlights the critical role of the proposed stage-wise framework in enhancing bug-fixing performance. Figure~\ref{rq2b} depicts an example of bug-fixing from the augmented BFP benchmark. Among the different ablation models (where T is short for $\mathbf{ChatGPT_{Tester}}$, D is short for $\mathbf{ChatGPT_{Developer}}$ and R is short for $\mathbf{ChatGPT_{Reviewer}}$), only \approach (i.e., $\mathbf{ChatGPT_{TDR}}$) manages to successfully patch the bug. The provided \colorbox{light-green}{\color{red}{commit message}} suggests that the root cause of the bug lies in a lack of control over the internal list, thus necessitating a patch that addresses the issue of duplicate states being added. As shown in Figure~\ref{rq2b}, it is evident that the ablation models without the $\mathbf{ChatGPT_{Reviewer}}$ component produce incorrect patches that either do not include any changes to the buggy hunk or fail to resolve the duplicate issue. However, with the guidance provided by the \colorbox{light-green}{\color{red}{commit message}} as the fixing goal during the patch verification stage, \approach is able to generate a correct candidate patch that matches the human-written ground truth.

\begin{table}[tbp]
\centering
    \caption{The Effect of Different Commit Message Types on Bug Fixing.}
    \label{rq2e}
    \resizebox{\textwidth}{!}{
    \begin{tabular}{lllllllllllll}
    \toprule
    \multirow{3.5}{*}{\textbf{{Model}}} & \multicolumn{3}{c}{\textbf{{Why and What}}} & \multicolumn{3}{c}{\textbf{{No Why}}} & \multicolumn{3}{c}{\textbf{{No What}}} & \multicolumn{3}{c}{\textbf{{Neither Why nor What}}} \\
    \cmidrule[0.5pt](rl){2-4}\cmidrule[0.5pt](rl){5-7}\cmidrule[0.5pt](rl){8-10}\cmidrule[0.5pt](rl){11-13}
    & \multicolumn{3}{c}{{1043 bugs}} & \multicolumn{3}{c}{{1541 bugs}} & \multicolumn{3}{c}{{314 bugs}} & \multicolumn{3}{c}{{214 bugs}} \\
    & \multicolumn{1}{c}{{$k=1$}} & \multicolumn{1}{c}{{$k=3$}} & \multicolumn{1}{c}{{$k=5$}} & \multicolumn{1}{c}{{$k=1$}} & \multicolumn{1}{c}{{$k=3$}} & \multicolumn{1}{c}{{$k=5$}} & \multicolumn{1}{c}{{$k=1$}} & \multicolumn{1}{c}{{$k=3$}} & \multicolumn{1}{c}{{$k=5$}} & \multicolumn{1}{c}{{$k=1$}} & \multicolumn{1}{c}{{$k=3$}} & \multicolumn{1}{c}{{$k=5$}} \\
    \midrule
    {ChatGPT} & {119} & {134} & {188} & {262} & {324} & {388} & {20} & {33} & {40} & {54} & {64} & {77} \\
    {ChatGPT w/ \texttt{CM}} & {189 \textbf{\color{red}(+70)}} & {246 \textbf{\color{red}(+112)}} & {277 \textbf{\color{red}(+89)}} & {316 \textbf{\color{red}(+54)}} & {405 \textbf{\color{red}(+81)}} & {449 \textbf{\color{red}(+61)}} & {28 \textbf{\color{red}(+8)}} & {41 \textbf{\color{red}(+8)}} & {45 \textbf{\color{red}(+5)}} & {54 \textbf{\color{red}(+0)}} & {65 \textbf{\color{red}(+1)}} & {78 \textbf{\color{red}(+1)}} \\
    \midrule
    {GPT-4} & {177} & {228} & {243} & {350} & {418} & {445} & {32} & {38} & {39} & {62} & {76} & {82} \\
    {GPT-4 w/ \texttt{CM}} & {265 \textbf{\color{red}(+88)}} & {292 \textbf{\color{red}(+64)}} & {315 \textbf{\color{red}(+72)}} & {428 \textbf{\color{red}(+78)}} & {483 \textbf{\color{red}(+65)}} & {521 \textbf{\color{red}(+76)}} & {45 \textbf{\color{red}(+13)}} & {55 \textbf{\color{red}(+17)}} & {58 \textbf{\color{red}(+19)}} & {70 \textbf{\color{red}(+8)}} & {84 \textbf{\color{red}(+8)}} & {90 \textbf{\color{red}(+8)}} \\
    \bottomrule
    \end{tabular}}
\end{table}

\begin{figure}[tbp]
  \centering
  \includegraphics[width=\textwidth]{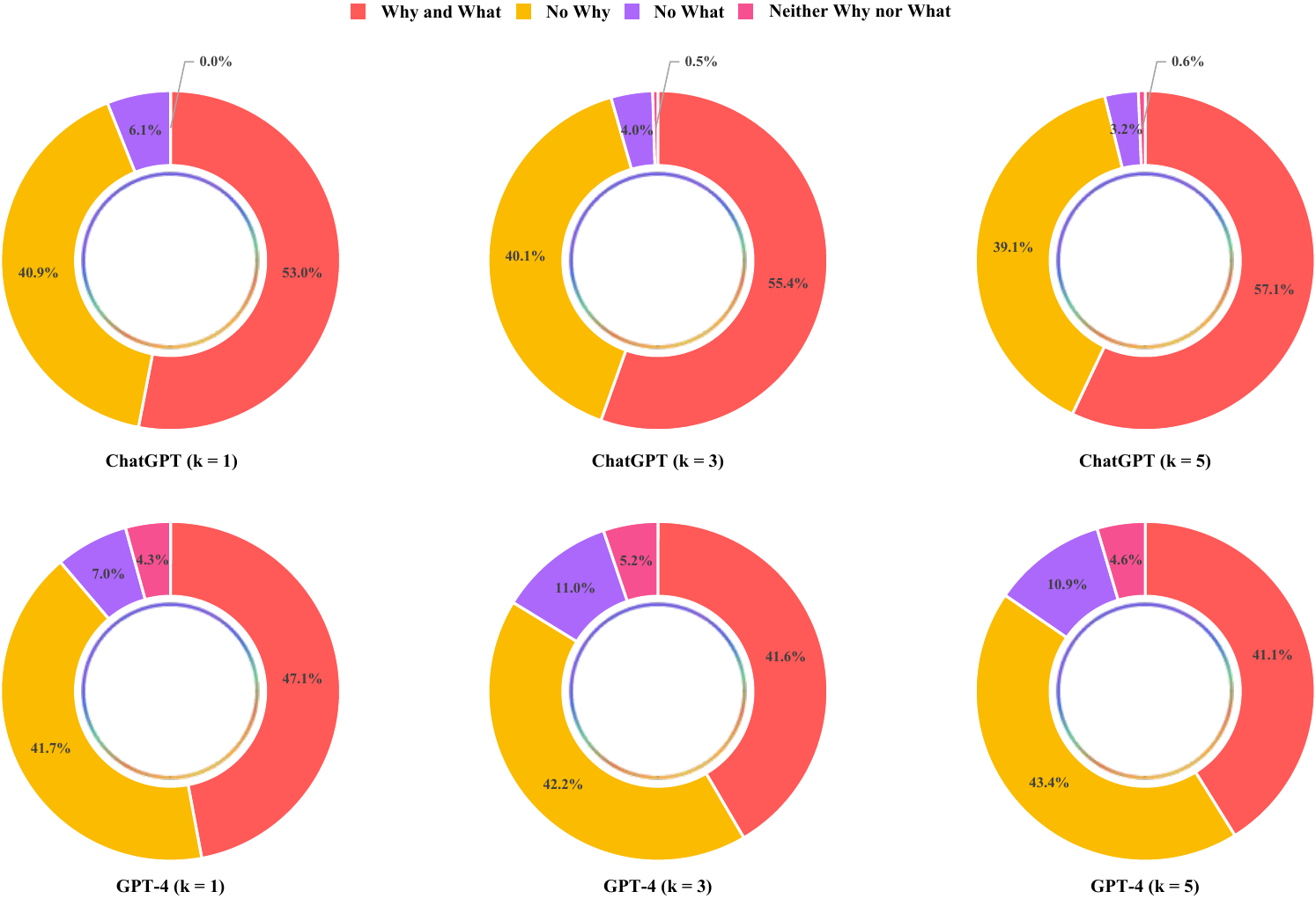}
  \caption{The Contribution of Different Commit Message Types to Improvements in Bug-Fixing Performance.}
  \label{rq2f}
  \Description{}
\end{figure}

\subsubsection{The Impact of Different Commit Message Types}

We further investigate the influence of commit message quality on bug-fixing performance. Following the approach outlined in existing research \cite{tian2022makes}, we categorize commit messages into four types based on the presence of \textbf{\textit{Why}} (i.e., justification for the commit change) and \textbf{\textit{What}} (i.e., summary of the commit change) information. Table~\ref{rq2e} provides a comparative analysis of two treatments: with and without the commit message (abbreviated as \texttt{CM}), across the four commit message types (\textit{Why and What}, \textit{No Why}, \textit{No What}, and \textit{Neither Why nor What}) for ChatGPT and GPT-4. The comparison examines the number of correct patches generated for each commit message type at $k=1$, $k=3$, and $k=5$. Our findings indicate that all four types have a non-negative effect on performance improvements, as they contribute to generating more correct patches. Furthermore, Figure~\ref{rq2f} illustrates the contributions of different commit message types to bug-fixing performance improvements. We observe that informative commit message types (\textit{Why and What}, \textit{No Why}, and \textit{No What}) contribute significantly more to performance gains compared to the non-informative type (\textit{Neither Why nor What}). This result is expected, as \textit{Neither Why nor What} type commit messages lack the necessary context, preventing the LLMs from reasoning accurately to fix the bug. On average, the three informative types account for over 95\% of the total contributions. Notably, for commit messages that contain only high-level information (i.e., \textit{No What}), the more advanced LLM GPT-4 achieves greater performance improvements than ChatGPT, highlighting its superior reasoning capabilities. Since the \textit{No What} type commit messages lack specific details about the steps taken to fix the bug, it forces the LLMs to rely on the justification for the commit change to infer how to resolve the bug, increasing the difficulty required for solving the bug-fixing task.

\subsubsection{The Impact of Interaction Turns}

\begin{figure}[htbp]
  \centering
  \includegraphics[width=0.48\textwidth]{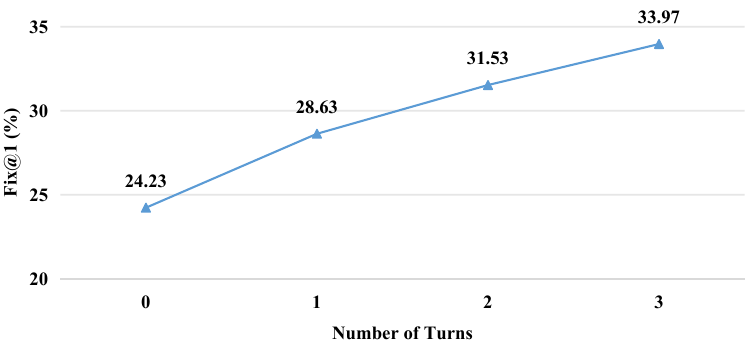}
  \caption{The Effect of Interaction Turns on Bug Fixing.}
  \label{rq2c}
  \Description{}
\end{figure}

\begin{figure}[tbp]
  \centering
  \includegraphics[width=\textwidth]{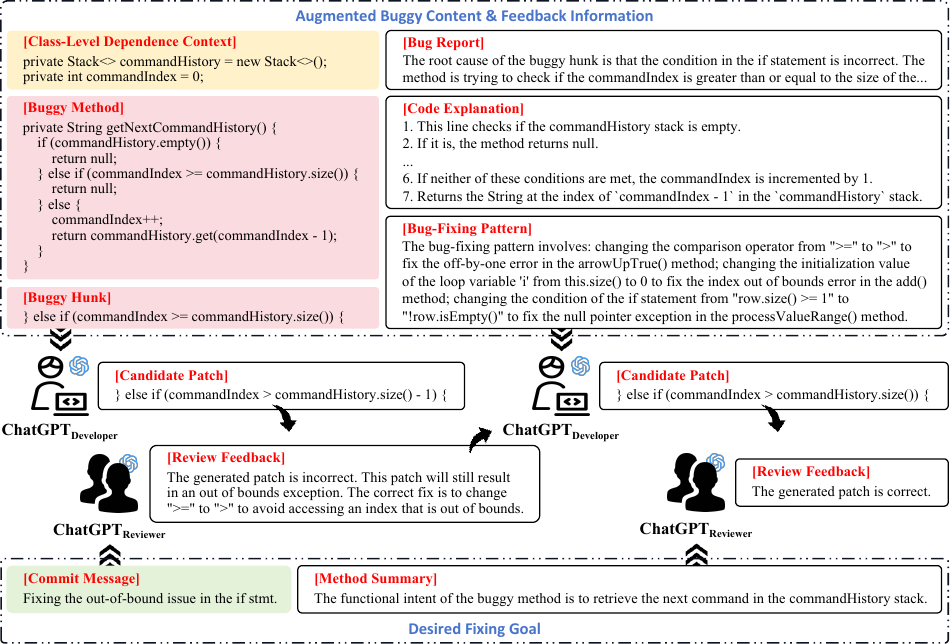}
  \caption{An Example of the Interactive Process between $\mathbf{ChatGPT_{Reviewer}}$ and $\mathbf{ChatGPT_{Developer}}$.}
  \label{rq2d}
  \Description{}
\end{figure}

In order to assess the impact of the interaction between $\mathbf{ChatGPT_{Reviewer}}$ and $\mathbf{ChatGPT_{Developer}}$, we control the number of interaction turns and illustrate the \textbf{Fix@1} results in Figure~\ref{rq2c}. When the number of interaction turns is set to zero, it signifies a complete absence of interaction between $\mathbf{ChatGPT_{Reviewer}}$ and $\mathbf{ChatGPT_{Developer}}$. Consequently, the candidate patch generated by $\mathbf{ChatGPT_{Developer}}$ receives no feedback message from $\mathbf{ChatGPT_{Reviewer}}$, leading to an outcome identical to that of the ablation model $\mathbf{ChatGPT_{TD}}$. It is noteworthy that the most significant improvement arises from the first interaction turn. Specifically, a single interaction turn with the $\mathbf{ChatGPT_{Reviewer}}$ component leads to an approximately 4.4\% enhancement in terms of \textbf{Fix@1} over the $\mathbf{ChatGPT_{TD}}$ model. As the number of interaction turns continues to increase beyond the initial round, the magnitude of improvements tends to diminish. Nonetheless, our observations indicate a consistent enhancement, suggesting that additional interactions still contribute to the capability to resolve more complex bugs. Figure~\ref{rq2d} illustrates an exemplary interactive process between $\mathbf{ChatGPT_{Reviewer}}$ and $\mathbf{ChatGPT_{Developer}}$ during the stages of patch generation and patch verification. In this scenario, $\mathbf{ChatGPT_{Reviewer}}$ determines that the \fcolorbox{black}{white}{\parbox{.135\linewidth}{\color{red}{candidate patch}}} generated by $\mathbf{ChatGPT_{Developer}}$ is incorrect based on the programmer's intent as summarized in the \colorbox{light-green}{\color{red}{commit message}}. Consequently, $\mathbf{ChatGPT_{Developer}}$ is required to generate a new patch while taking into account the \fcolorbox{black}{white}{\parbox{.135\linewidth}{\color{red}{review feedback}}}. As shown in Figure~\ref{rq2d}, \approach successfully generates the correct patch through this collaborative effort.

\begin{tcolorbox}[size=title]
    \textbf{Answer to RQ2}: To sum up, all components of \approach significantly improve the fixing performance. Regarding \textbf{Fix@1}, the addition of $\mathbf{ChatGPT_{Tester}}$ enhances ChatGPT by 2.86\%. Furthermore, the introduction of $\mathbf{ChatGPT_{Developer}}$ leads to additional improvements of 6.75\%, while the incorporation of $\mathbf{ChatGPT_{Reviewer}}$ results in continuous enhancements of 9.74\%.
\end{tcolorbox}

\subsection{Answering RQ3}

To answer this question, we conduct an extensive evaluation of the generalizability of \approach across four widely used APR benchmarks and five popular open-source LLM baselines. This rigorous analysis offers a comprehensive assessment of the effectiveness and robustness of \approach.

\begin{table}[tbp]
\centering
    \caption{Comparison of \approach on Four APR Benchmarks against 17 Baselines.}
    \label{rq3a}
    \begin{tabular}{lrrrrr}
    \toprule
    \multicolumn{1}{c}{\multirow{4}{*}{\textbf{Model}}} & \multicolumn{2}{c}{\textbf{\# of \textit{Exact Match} Patches}} & \multicolumn{3}{c}{\textbf{\# of \textit{Correct} Patches}} \\
    \cmidrule[0.5pt](rl){2-3}\cmidrule[0.5pt](rl){4-6}
    & \multicolumn{1}{c}{\textbf{Bugs.jar}} & \multicolumn{1}{c}{\textbf{Bears}} & \multicolumn{1}{c}{\multirow{2.5}{*}{\textbf{Time / \# Patch}}} & \multicolumn{1}{c}{\textbf{Defects4J}} & \multicolumn{1}{c}{\textbf{QuixBugs}} \\
    \cmidrule[0.5pt](rl){2-2}\cmidrule[0.5pt](rl){3-3}\cmidrule[0.5pt](rl){5-5}\cmidrule[0.5pt](rl){6-6}
    & \multicolumn{1}{c}{1000 bugs} & \multicolumn{1}{c}{119 bugs} & & \multicolumn{1}{c}{313 bugs} & \multicolumn{1}{c}{37 bugs} \\
    \midrule
    {TBar} & - & - & {3 Hour} & {63} & {-} \\
    \midrule
    {CoCoNut} & 66 & 16 & {$1000$} & {47} & {13} \\
    {{\sc {SequenceR}}\xspace} & 99 & 14 & {$300$} & {41} & {15} \\
    {Recoder} & 61 & 1 & {5 Hour} & {60} & {17} \\
    {CURE} & {-} & {-} & {$10000$} & {59} & {25} \\
    {RewardRepair} & 103 & 8 & {$200$} & {72} & {20} \\
    {SelfAPR} & {-} & {-} & {-} & {86} & {-} \\
    {KNOD} & {-} & {-} & {$1000$} & {102} & {25} \\
    \midrule
    {CodeBERT} & 111 & 12 & {-} & {-} & {-} \\
    {GraphCodeBERT} & 115 & 12 & {-} & {-} & {-} \\
    {CodeT5} & 150 & 16 & {-} & {-} & {-} \\
    {UniXcoder} & 164 & 22 & {-} & {-} & {-} \\
    \midrule
    AlphaRepair & - & - & {$\leq 5000$} & {92} & {28} \\
    {Repilot} & {-} & {-} & {$\leq 5000$} & {116} & {-} \\
    ChatRepair & - & - & {$\leq 200$} & {142} & \textbf{{36}} \\
    ThinkRepair & - & - & {$\leq 125$} & {155} & \textbf{{36}} \\
    {RepairAgent} & {-} & {-} & {Avg. of $117$} & {124} & {-} \\
    \midrule
    \approach & \textbf{180 {\color{red}(9.76\% $\uparrow$)}} & \textbf{28 {\color{red}(27.27\% $\uparrow$)}} & {$\leq 100$} & \textbf{{169} {\color{red}(9.03\% $\uparrow$)}} & {35} {\color{green}(2.86\% $\downarrow$)} \\
    \bottomrule
    \end{tabular}
\end{table}

\subsubsection{Generalizability Evaluation on APR Benchmarks}
\label{rq3-1}

We select single-hunk bugs from four additional APR benchmarks for performance evaluation: Bugs.jar \cite{saha2018bugsjar} (1000 bugs), Bears \cite{delfim2019bears} (119 bugs), Defects4J \cite{just2014defects4j} (313 bugs), and QuixBugs \cite{lin2017quixbugs} (37 bugs). Specifically, we compare \approach to 17 APR baselines, which include template-based, neural-based, pre-trained code language model-based, and LLM-based approaches. For template-based APR, we use the state-of-the-art approach TBar \cite{liu2019tbar} with perfect fault localization. For neural-based APR, we select seven recently published approaches: CoCoNut \cite{lutellier2020coconut}, {\sc SequenceR\xspace} \cite{chen2021sequencer}, Recoder \cite{zhu2021syntax}, CURE \cite{jiang2021cure}, RewardRepair \cite{ye2022neural}, SelfAPR \cite{ye2022selfapr}, and KNOD \cite{jiang2023knod}. We further include four pre-trained code language models: CodeBERT \cite{feng2020codebert}, GraphCodeBERT \cite{guo2021graphcodebert}, CodeT5 \cite{wang2021codet5}, and UniXcoder \cite{guo2022unixcoder}. Given that the BFP benchmark lacks test suites, we evaluate \approach by comparing it to {five} recent LLM-based APR approaches in this RQ: AlphaRepair \cite{xia2022less}, {Repilot \cite{wei2023copiloting}}, ChatRepair \cite{xia2024automated}, ThinkRepair \cite{yin2024thinkrepair}, {and RepairAgent \cite{bouzenia2024repairagent}}. These LLM-based approaches are evaluated on the Defects4J and QuixBugs benchmarks, both of which include test suites.

\textbf{Experimental Settings.} For Bugs.jar and Bears, we use the \textit{exact match} metric to evaluate the correctness of each generated candidate patch. This metric is preferable for these benchmarks due to their lack of test suites, which helps avoid human bias and minimizes the manual effort required for patch validation \cite{ye2021automated}. In these experiments, \approach generates the top-1 candidate patch for each bug, whereas the selected baselines typically generate a larger number of candidates (more than 100), as demonstrated in related studies \cite{zhong2022standupnpr,zhong2024benchmarking}. This paper reports the number of \textit{exact match} patches within the top-10 candidates generated by each baseline \cite{lutellier2020coconut,chen2021sequencer,zhu2021syntax,ye2022neural,feng2020codebert,guo2021graphcodebert,wang2021codet5,guo2022unixcoder}, consistent with recent findings \cite{noller2022trust}, which suggest that most developers are willing to review no more than 10 patches. For Defects4J and QuixBugs, we adopt the widely-used \textit{test-passing} metric for patch assessment, following previous studies \cite{lutellier2020coconut,chen2021sequencer,zhu2021syntax,jiang2021cure,ye2022neural,ye2022selfapr,jiang2023knod,xia2022less,wei2023copiloting,xia2024automated,yin2024thinkrepair,bouzenia2024repairagent}. A patch is considered \textit{plausible} if it passes all test cases. All \textit{plausible} patches are subsequently manually reviewed by comparing them to the developer-written ground truth. A patch is deemed \textit{correct} if it is either 1) identical to the developer-written patch or 2) semantically equivalent to the developer-written patch. In our experiments, \approach samples a small-scale set of candidate patches (at most 100 fixing attempts), fewer than those generated by all compared baselines, with a sampling temperature of 1 to ensure response diversity. This paper reports the number of \textit{correct} patches generated by \approach within these 100 attempts. In evaluating the results of corresponding baselines, we carefully compare the bug IDs correctly reported in the original papers with the set of single-hunk bugs selected from the Defects4J and QuixBugs benchmarks utilized in this paper.

\textbf{Evaluation Results.} Table~\ref{rq3a} illustrates the number of bugs successfully fixed by \approach and each baseline across different benchmarks in the context of fixing sing-hunk bugs. The \textbf{Time / \# Patch} column indicates the maximum time allowed for fixing a bug or the maximum number of fixing attempts permitted for a model to generate before a \textit{plausible} patch is obtained. A ``-'' indicates that no results have been reported in the corresponding papers. The best performance for each benchmark is highlighted in bold. As observed, \approach demonstrates a substantial performance advantage over the baseline approaches across three benchmarks (except QuixBugs). Specifically, \approach surpasses the best baseline by 9.76\% in the Bugs.jar benchmark (compared to UniXcoder), 27.27\% in the Bears benchmark (compared to UniXcoder), and 9.03\% in the Defects4J benchmark (compared to ThinkRepair). For the QuixBugs benchmark, the two LLM-based APR approaches, namely ChatRepair and ThinkRepair, achieve superior performance, surpassing \approach by 2.86\%. This difference in performance may be attributed to two primary factors: first, these approaches leverage error message feedback obtained from the validation process using test suites; second, they sample a significantly larger number of candidate patches (over 100) for fixing a corresponding bug compared to \approach. Consequently, the number of bugs fixed by \approach on the QuixBugs benchmark is lower than that of ChatRepair and ThinkRepair.

\begin{figure}[tbp]
  \centering
  \includegraphics[width=0.98\textwidth]{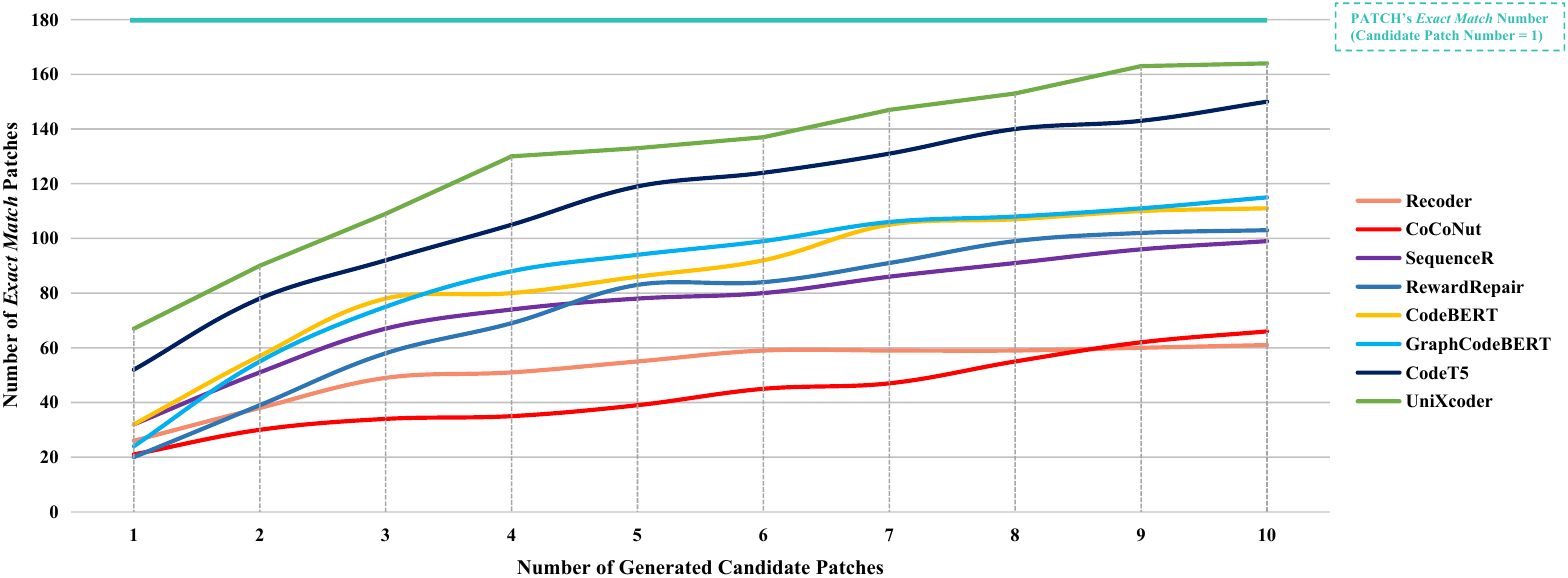}
  \caption{The Number of Correct Patches Generated by Each Baseline on Bugs.jar under Different Candidate Numbers.}
  \label{rq3b}
  \Description{}
\end{figure}

\textbf{Impact of the Candidate Number.} We investigate the number of correct patches generated by each baseline on the Bugs.jar benchmark under different candidate numbers. As depicted in Figure~\ref{rq3b}, a large candidate set clearly increases the likelihood of containing the \textit{exact match} patch. Notably, all baselines demonstrated improved bug-fixing performance with the increasing number of candidates. Nevertheless, it is noteworthy that \approach consistently outperforms the baseline approaches when applied to greedy decoding (i.e., generating top-1 candidate patch for each bug).

\begin{figure}[tbp]
  \centering
  \includegraphics[width=\textwidth]{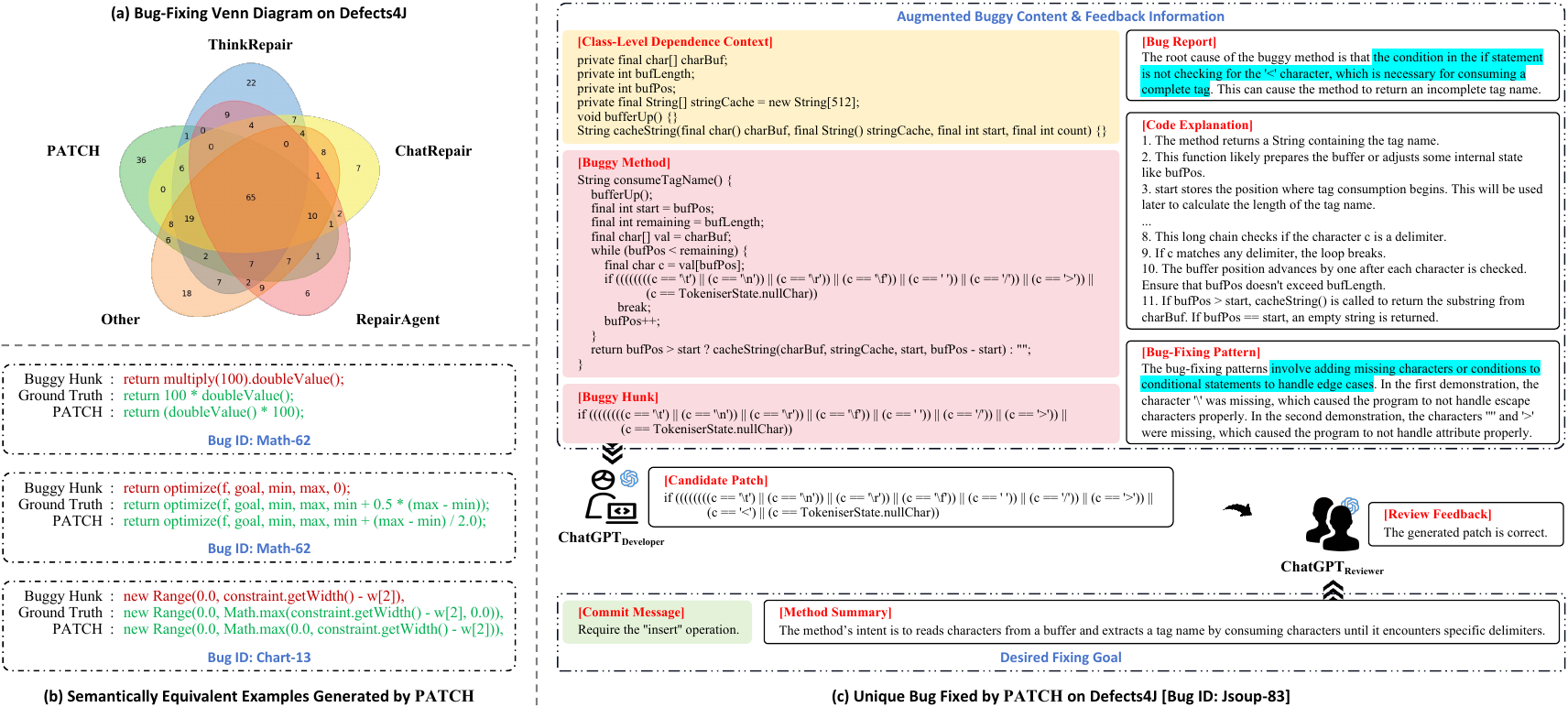}
  \caption{Examples of the Bug-Fixing Venn Diagram, Equivalent Patches, and a Uniquely Fixed Bug by \approach on Defects4J.}
  \label{rq3d}
  \Description{}
\end{figure}

\textbf{Case Study on Defects4J.} We present a Venn diagram (shown in Figure~\ref{rq3d}(a)) to further illustrate the bug-fixing performance across different approaches on Defects4J. For clarity, we highlight the top three best-performing baselines (i.e., ThinkRepair, ChatRepair, and RepairAgent) based on the number of correctly fixed bugs, and group all distinct correctly fixed bugs by the remaining baselines into an ``Other'' category for easier comparison. Two observations can be drawn from Figure~\ref{rq3d}(a): \ding{172} Individual approaches exhibit varying bug-fixing capabilities, with each being able to fix some specific bugs that others cannot. This suggests a degree of complementary performance among the approaches. \ding{173} Overall, \approach generates the highest number of unique bug fixes (i.e., 36) that other baselines are unable to resolve. Figure~\ref{rq3d}(b) showcases three examples generated by \approach that, while not exact matches, are semantically equivalent to the developer-written ground truth. For instance, in \textbf{Math-62}, the developer uses ``$0.5 * (max - min)$'' for the calculation, while \approach employs the equivalent operation, i.e., ``$(max - min) / 2.0$''. Additionally, we provide a bug (\textbf{Jsoup-83}) from Defects4J, depicted in Figure~\ref{rq3d}(c), that is uniquely fixed by \approach. Specifically, we use the required edit operation (i.e., insert) to fix the bug as the \colorbox{light-green}{\color{red}{commit message}} for prompting \approach. It is reasonable to provide only the high-level description of the bug type without detailing the resolution process. As observed, \approach benefits from the \colorbox{light-blue3}{feedback information} within the \fcolorbox{black}{white}{\parbox{.09\linewidth}{\color{red}{bug report}}} and \fcolorbox{black}{white}{\parbox{.16\linewidth}{\color{red}{bug-fixing patterns}}}. \approach correctly introduces the necessary condition (i.e., ``$<$'') in the if statement to handle edge cases, ensuring the validity of consuming a complete tag. In this case, both the root cause identified by $\mathbf{ChatGPT_{Tester}}$ and the patterns derived from the retrieved similar demonstrations, as summarized by $\mathbf{ChatGPT_{Developer}}$, contribute to the successful resolution of \textbf{Jsoup-83}, highlighting the collaborative capabilities of \approach.

\begin{table}[tbp]
    \centering
    \caption{Generalizability of \approach on Different LLMs.}
    \label{rq3c}
    \resizebox{\textwidth}{!}{
    \begin{tabular}{llllll}
        \toprule
        \multicolumn{1}{c}{\textbf{Model}} & \textbf{Fix@1 (\%) $\uparrow$} & \textbf{Simple Delete} & \textbf{Simple Insert} & \textbf{Simple Replace} & \textbf{Mixed} \\
        \midrule
        DeepSeek-Coder-1.3B & 8.42 & 28 & 14 & 112 & 108 \\
        DeepSeek-Coder-1.3B w/ \approach & 9.19 \textbf{\color{red}(9.14\% $\uparrow$)} & 29 \textbf{\color{red}(+1)} & 17 \textbf{\color{red}(+3)} & 120 \textbf{\color{red}(+8)} & 120 \textbf{\color{red}(+12)} \\
        \midrule
        CodeGeeX2-6B & 10.93 & 23 & 51 & 129 & 137 \\
        CodeGeeX2-6B w/ \approach & 13.95 \textbf{\color{red}(27.63\% $\uparrow$)} & 25 \textbf{\color{red}(+2)} & 57 \textbf{\color{red}(+6)} & 192 \textbf{\color{red}(+63)} & 160 \textbf{\color{red}(+23)} \\
        \midrule
        Mistral-7B & 9.38 & 29 & 16 & 124 & 123 \\
        Mistral-7B w/ \approach & 15.68 \textbf{\color{red}(67.16\% $\uparrow$)} & 39 \textbf{\color{red}(+10)} & 28 \textbf{\color{red}(+12)} & 244 \textbf{\color{red}(+120)} & 177 \textbf{\color{red}(+54)} \\
        \midrule
        CodeLLaMA-7B & 9.45 & 29 & 13 & 136 & 116 \\
        CodeLLaMA-7B w/ \approach & 15.87 \textbf{\color{red}(67.94\% $\uparrow$)} & 35 \textbf{\color{red}(+6)} & 25 \textbf{\color{red}(+12)} & 249 \textbf{\color{red}(+113)} & 185 \textbf{\color{red}(+69)} \\
        \midrule
        CodeLLaMA-13B & 9.80 & 28 & 14 & 137 & 126 \\
        CodeLLaMA-13B w/ \approach & 18.83 \textbf{\color{red}(92.14\% $\uparrow$)} & 38 \textbf{\color{red}(+10)} & 34 \textbf{\color{red}(+20)} & 287 \textbf{\color{red}(+150)} & 227 \textbf{\color{red}(+101)} \\
        \bottomrule
    \end{tabular}}
\end{table}

\subsubsection{Generalizability Evaluation on Open-Source LLMs}

We further employ \approach to five open-source LLMs, including DeepSeek-Coder-1.3B, CodeGeeX2-6B, Mistral-7B, CodeLLaMA-7B, and CodeLLaMA-13B, all of which support interactive dialogues. Specifically, we conduct ablation experiments on each model individually to investigate the impact of interactively querying the corresponding LLM with devised prompts as specified in \approach. To ensure the comparison fairness, the parameter configurations align with those described in Section~\ref{implementation}. Table~\ref{rq3c} presents the comparison results using the \textbf{Fix@1} metric, alongside the number of correct patches generated for each bug type. Each model's results are depicted in two lines: the first line shows the results when the corresponding LLM directly utilizes the buggy code and a few bug-fixing demonstrations as the input prompt to resolve the given bug, and the second line shows the results when integrated with \approach. From Table~\ref{rq3c}, we derive several key insights. \textbf{\ding{172}} \textbf{Integrating \approach as a complementary plug-in enhances the bug-fixing performance across all LLMs.} The observed performance improvements range from 9.14\% to 92.14\%. \textbf{\ding{173}} \textbf{The performance gains exhibit a marked increase with the number of parameters exceeding 7B.} For instance, the performance gain observed with the DeepSeek-Coder-1.3B model is substantially smaller than that of the CodeLLaMA-13B model. \textbf{\ding{174}} \textbf{The number of correct patches generated using \approach is consistently higher across all types of bugs}, particularly for complex bug types such as \textbf{Simple Replace} and \textbf{Mixed}. In summary, \approach effectively harnesses the latent intelligence of the LLMs in an interactive and collaborative manner, thereby improving their performance in the bug-fixing task.

\begin{tcolorbox}[size=title]
    \textbf{Answer to RQ3}: Since \approach does not rely on fine-tuning with specific bug-fixing datasets, it demonstrates a lower susceptibility to generalizability issues. Consequently, \approach outperforms traditional approaches across various APR benchmarks. Looking ahead, \approach holds the potential to be seamlessly integrated with additional LLMs in a plug-and-play manner.
\end{tcolorbox}

\section{Discussion}
\label{dis}

\subsection{Integration with Fine-Tuned LLMs}

\begin{figure}[htbp]
  \centering
  \includegraphics[width=0.48\textwidth]{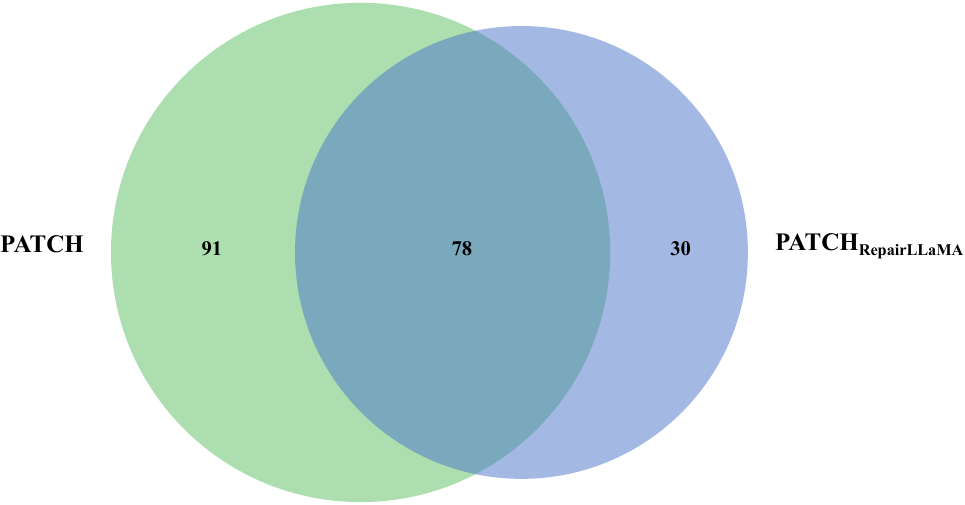}
  \caption{Bug-fixing Venn Diagram between \approach and \approach$_{\mathbf{RepairLLaMA}}$ on Defects4J.}
  \label{threat1}
  \Description{}
\end{figure}

We conduct a preliminary experiment by replacing the $\mathbf{ChatGPT_{Developer}}$ component in the patch generation stage with the fine-tuned LLM RepairLLaMA \cite{silva2023repairllama} for APR, using the Defects4J benchmark. Figure~\ref{threat1} illustrates the intersection of correct bug fixes between \approach and the variant model \approach$_{\mathbf{RepairLLaMA}}$. The experimental results indicates that \approach, using vanilla ChatGPT, outperforms the variant using the fine-tuned LLaMA model. The performance decline can be attributed to two possible factors: 1) LLaMA has fewer parameters compared to ChatGPT, and 2) the input length for RepairLLaMA is constrained to 1024 tokens, which may cause certain bug examples in Defects4J, specifically those with longer buggy contexts and feedback information, to exceed this limit. Consequently, the absence of contextual information may result from the utilization of truncation strategies. Nevertheless, \approach$_{\mathbf{RepairLLaMA}}$ is still able to fix 30 bugs that \approach could not address. Exploring more effective ways to integrate fine-tuned LLMs into \approach will also remain as our future work.

\subsection{Multi-Hunk Bug-Fixing Scenario}
\label{multi}

\begin{figure}[htbp]
  \centering
  \includegraphics[width=\textwidth]{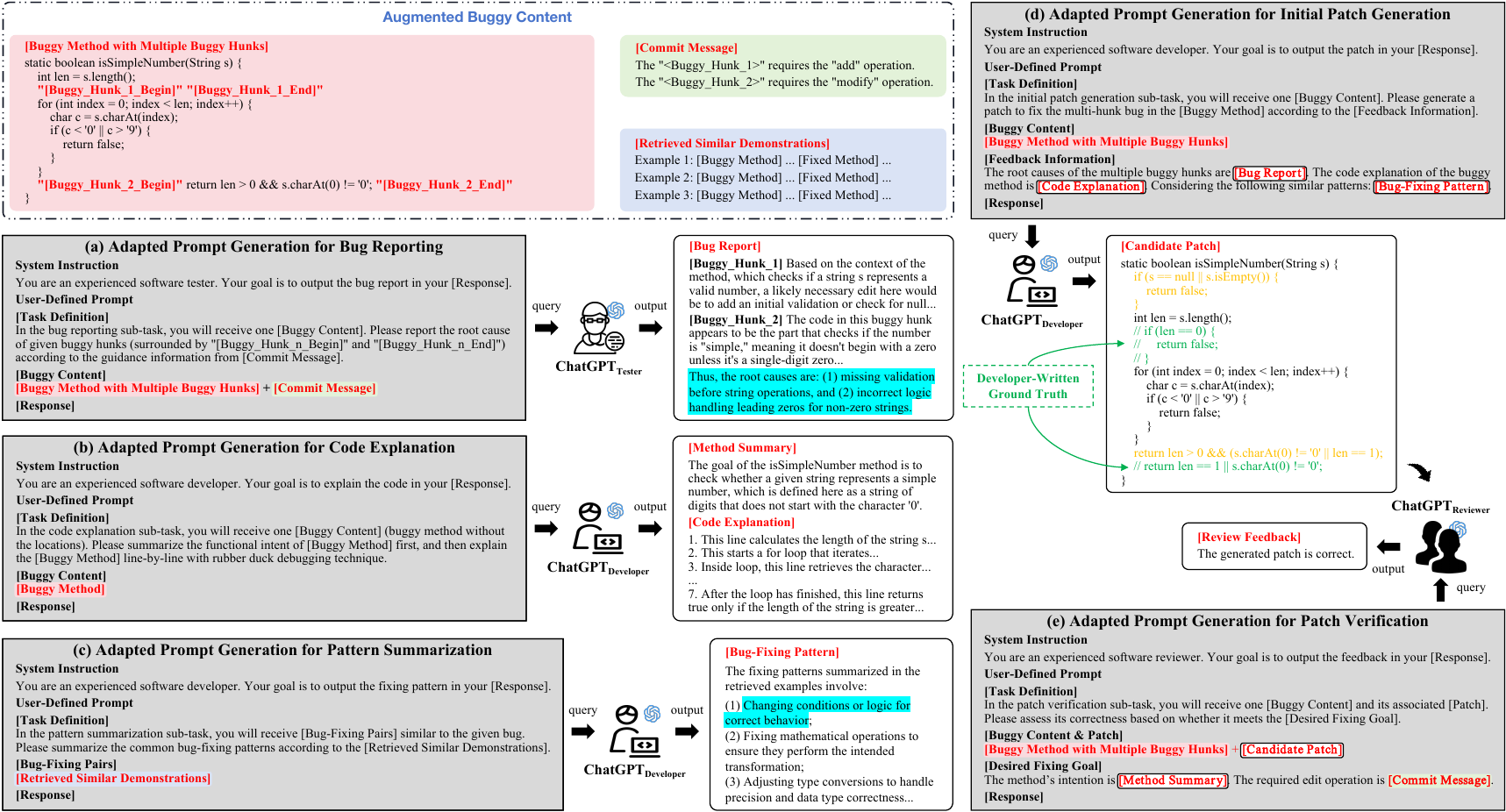}
  \caption{An Example of Adaptation Prompts Used by \approach for Fixing the Multi-Hunk Bug \textbf{Closure-128} from Defects4J.}
  \label{threat2}
  \Description{}
\end{figure}

Figure~\ref{threat2} presents an example of fixing the multi-hunk bug \textbf{Closure-128} from Defects4J using \approach. Figure~\ref{threat2}(a)-(e) illustrate the adapted prompts for the single-function multi-hunk fixing scenario at different bug-fixing stages. In the adapted framework, we modify the code representation of the \colorbox{light-red}{\color{red}{buggy method}} by enclosing each \colorbox{light-red}{\color{red}{buggy hunk}} with specific identifiers (i.e., {\color{red}\textbf{"[Buggy\_Hunk\_n\_Begin]"}} and {\color{red}\textbf{"[Buggy\_Hunk\_n\_End]"}}). Accordingly, \approach generates the complete, fixed method as a candidate patch. As shown in Figure~\ref{threat2}(d), \approach successfully fixes \textbf{Closure-128} in the initial patch generation stage, producing a patch is semantically equivalent to the developer-written one. This is achieved through leveraging \colorbox{light-blue3}{feedback information} from earlier stages. Specifically, $\mathbf{ChatGPT_{Tester}}$ identifies the root causes of the two \colorbox{light-red}{\color{red}{buggy hunks}}: \textbf{Buggy\_Hunk\_1} is caused by \textit{missing validation before string operations}, and \textbf{Buggy\_Hunk\_2} results from \textit{incorrect logic handling leading zeros for non-zero strings}. Subsequently, $\mathbf{ChatGPT_{Developer}}$ retrieves a similar bug-fixing example involving the patten of \textit{changing conditions or logic for correct behavior}. In the patch generation step, \approach addresses the first \colorbox{light-red}{\color{red}{buggy hunk}} by adding validation for null or empty string, and fixes the second \colorbox{light-red}{\color{red}{buggy hunk}} by adjusting the logic for handling leading zeros. Overall, \approach can be adapted to multi-hunk bug-fixing scenarios by adjusting the content of corresponding prompts to accommodate the specific structure of these bugs.

\subsection{Threats to Validity}

In this subsection, we outline the main threats to the validity of \approach, as detailed below:
\begin{itemize}
	\item \textbf{External threat}: The primary threats to external validity in this paper pertain to the quality of the selected experimental subjects and the use of the commit message. The extent to which the improvements achieved by \approach are applicable to other bug-fixing benchmarks remains uncertain. To mitigate this concern, we have utilized the mainstream benchmark BFP, consistent with prior studies \cite{tufano2019empirical,tang2021grammar,wang2021codet5,zhong2022standupnpr}, and supplemented the evaluation with four additional APR benchmarks to enhance the diversity of the evaluation. Furthermore, programmers typically write commit messages after fixing buggy code. In this paper, we assumed that programmers write these messages before bug fixing, potentially limiting \approach's real-world applicability. Nevertheless, we view \approach as a proof-of-concept. The empirical results indicate that utilizing commit messages as input can aid LLMs in comprehending programmers' reasoning processes. These messages offer valuable insights into the programmer's intent during the fix, thereby empowering LLMs with guided context. In addition, the retrieved bug-fixing pairs during the pattern summarization stage are crucial components of \approach. Intuitively, when the retrieved code is less similar to the given bug, the performance of \approach may degrade. We apply a dynamic threshold (described in Section~\ref{patsum}) to ensure the quality of the retrieval. Empirical evidence also suggests that code reuse can reach up to 80\% in real-world projects \cite{nguyen2013study}. Therefore, we believe that retrieving relevant examples for summarizing bug-fixing patterns is highly feasible in practical development scenarios.
	\item \textbf{Internal threat}: LLMs exhibit sensitivity to prompts and hyper-parameters, especially concerning the number of task examples and natural language instructions, which can significantly affect their performance. To minimize this variability, we employ consistent prompts and hyper-parameters for \approach and the LLM baselines. We refrain from experimental tuning of the prompt design and hyper-parameters, opting instead for empirical settings. As a result, we recognize that there is potential for further performance improvements through additional tuning. Another potential threat involves the possibility that the BFP benchmark, comprising Java projects hosted on GitHub between 2011 and 2017, may have been included in the training data of ChatGPT, raising concerns about the data leakage issue. Since ChatGPT is a closed-source model, the exact composition of its training data remains unknown. Despite this limitation, \approach exhibits a significant improvement in bug-fixing performance compared to the base ChatGPT model, which employs the same underlying architecture. This enhancement suggests that the superior results achieved by \approach are not merely attributable to the model's memorization of its training data. Additionally, the equivalent patch examples shown in Figure~\ref{rq3d}(b) and Figure~\ref{threat2}(d) further supports our argument, as the Defects4J benchmark may be included in the training data of the GPT-series models \cite{lee2024github}. Furthermore, while occasional inaccuracies arise, LLMs effectively capture the nature of bugs and provide clear, coherent code explanations. To reduce the likelihood of generating incorrect patches, we initially input commit messages at the bug reporting stage to assist ChatGPT in understanding the root cause of the given bug. Subsequently, we introduce the patch verification stage for further quality refinement.
	\item \textbf{Construct threat}: This paper uses \textbf{Fix@1} to assess the correctness of the generated candidate patches. Although this metric does not fully capture human judgment, it offers a strict and objective measure, facilitating rapid and quantitative evaluation of the model's performance. For the evaluation of APR benchmarks with test suites in Section~\ref{rq3-1}, we adopt the widely-used \textit{test-passing} metric to ensure a fair comparison. Future work will include additional human evaluations to further validate the models. In addition, \approach specifically targets the resolution of single-hunk Java bugs utilizing ChatGPT. It is essential to note, however, that the components designed for \approach can be effectively applied to various programming languages, integrated with multiple LLMs, and utilized in a wider range of bug-fixing scenarios. In Section~\ref{multi}, we present a case study evaluating \approach in fixing single-function multi-hunk bugs through prompt adaptation. In future work, we aim to expand the evaluation scope of bug fixing to comprehensively assess the generalizability of \approach.
\end{itemize}

\section{Related Work}
\label{rel}

\subsection{Automatic Bug Fixing}

Over the last decade, automatic bug fixing has emerged as a promising research topic, garnering significant attention from both the SE and AI communities. Traditional approaches can be broadly divided into two categories: search-based \cite{goues2012genprog,le2016history,saha2019harnessing,ghanbari2019practical,liu2019tbar,koyuncu2020fixminer} and semantics-based \cite{mechtaev2016angelix,xuan2017nopol,le2017syntax,chen2021contract,afzal2021sosrepair}. Search-based approaches typically rely on predefined bug-fixing patterns mined from historical open-source software repositories to generate patches, whereas semantics-based approaches generate patches by solving repair constraints derived from test suite specifications.

With the rapid advancement of DL techniques, there has been an increasing focus on neural-based approaches \cite{zhong2022neural,zhang2024survey}, which have shown remarkable potential in enhancing bug-fixing performance. Unlike traditional bug-fixing approaches, learning-based techniques can automatically capture semantic relationships between parallel bug-fixing pairs, enabling the generation of more effective and context-aware patch solutions. However, candidate patches generated by pre-trained models are typically not evaluated against a test suite or subjected to other automated verification strategies. Consequently, these patches may encounter issues related to compilability. In contrast to existing learning-based studies that typically use only static source code as input, SelfAPR \cite{ye2022selfapr} extracts test execution diagnostics and encodes them into the input representation, guiding neural models in fixing specific bugs. Furthermore, while most existing approaches are designed to fix bugs at a single location, several multi-hunk bug-fixing methods have been proposed, utilizing either a divide-and-conquer strategy \cite{li2022dear} or an iterative fixing paradigm \cite{ye2024iter}.

Recently, researchers have explored the feasibility of employing LLMs for automatic bug fixing. LLMs have demonstrated the capability to directly generate correct patches based on the surrounding context, obviating the necessity for fine-tuning. Despite the unprecedented outcomes achieved by LLM-based approaches \cite{xia2023automated,prenner2022openai,jiang2023impact,sobania2023analysis}, these techniques primarily focus on the buggy code and treat the bug fixing process as a single-stage task, neglecting the interactive and collaborative nature inherent in bug resolution. Nevertheless, recent advancements have shifted towards multi-step approaches. For example, ChatRepair \cite{xia2024automated} employs a conversational-driven approach, iteratively querying the LLM based on relevant test failure information derived from previous fix attempts. Similarly, ThinkRepair \cite{yin2024thinkrepair} tackles bug fixing through a two-step process: first performing few-shot learning using retrieved examples, followed by automatic interactions with LLMs, supplemented by feedback from test results. In contrast, RepairAgent \cite{bouzenia2024repairagent} allows LLMs to interact with predefined tools that assist in the bug-fixing process. This paper introduces a stage-wise framework comprising multiple ChatGPT agents, each assigned to distinct stages within the bug management process using specific prompts. To the best of our knowledge, this is the first attempt to enhance the bug-fixing capabilities of LLMs through the guidance of programmer intent and the interactive simulation of collaborative behavior.

\subsection{Large Language Model}

Recent advancements in generative AI have led to a significant increase in the performance and widespread adoption of LLMs \cite{xin2023survey}. LLMs undergo unsupervised training using billions of open-source text/code tokens to achieve comprehensive language modeling. Due to the utilization of diverse data sources and their general design to acquire cross-domain knowledge, researchers can subsequently utilize LLMs for specific downstream tasks (e.g., improving the efficiency of programmers in writing, editing, and reviewing code \cite{li2023codeeditor,zhang2024deep,tufano2024code}) by providing tailored prompts or a few demonstrations of the task being solved as input \cite{liu2023pretrain}. Among LLMs, the GPT family developed by OpenAI \cite{brown2020language,chen2021evaluating} is particularly notable for its popularity and prowess. Additionally, numerous attempts have been made to reproduce similar open-source LLMs such as CodeGPT \cite{lu2021codexglue}, LLaMA \cite{touvron2023llama}, and others. Despite their robust performance, LLMs sometimes struggle to produce accurate results when faced with complex tasks. In response, researchers have proposed the use of advanced prompting techniques (e.g., Chain of Thoughts (CoT) \cite{wei2022chain} and Tree of Thoughts (ToT) \cite{yao2023tree}) to enhance the reasoning capability of LLMs in natural language processing tasks. These techniques involve a sequence of intermediate reasoning steps in natural language that culminate in the final output. More recently, researchers have proposed LLMs trained using reinforcement learning to better align with human preferences. Examples of such models include InstructGPT \cite{ouyang2022training} and ChatGPT \cite{openai2022chatgpt}, which are initially initialized from a pre-trained model on autoregressive generation and then fine-tuned using reinforcement learning from human feedback (RLHF) \cite{ziegler2019finetuning}. This fine-tuning process, which incorporates human preferences, has significantly enhanced the ability of these LLMs to comprehend input prompts and follow instructions to perform complex tasks \cite{bang2023multitask}. Notably, ChatGPT has achieved superior performance in various SE tasks \cite{dong2024self,sobania2023analysis}. The objective of this paper is to draw insights from effective bug management practices to enhance the capabilities of existing LLMs in the task of bug fixing. Our experimental results demonstrate that such alignment enables ChatGPT to interact and collaborate, significantly outperforming traditional LLMs.

\section{Conclusion and Future Work}
\label{con}

This paper introduces a stage-wise framework aimed at enhancing the bug-fixing capabilities of LLMs in an interactive manner. We explore the potential of ChatGPT by simulating the behavior of human programmers engaged in bug management. Specifically, we augment the BFP benchmark by providing contextual information to better guide LLMs in generating the correct patches. Moreover, we decompose the bug-fixing task into four distinct stages and employ three ChatGPT agents to collectively produce candidate patches for bug resolution using devised prompts. We conduct extensive experiments to demonstrate the effectiveness of \approach. We firmly believe that aligning the collaborative problem-solving skills of programmers with LLMs represents a pivotal stride towards intelligent SE research.

\begin{acks}
We would like to thank the reviewers for their insightful comments. This work was partially supported by the Strategic Priority Research Program of the Chinese Academy of Sciences (Grant Nos. XDA0320100 and XDA0320102), the National Natural Science Foundation of China (Grant Nos. 62192731, 62192733, and 62192730), the Major Program (JD) of Hubei Province (Grant No. 2023BAA024), the Major Project of ISCAS (Grant No. ISCAS-ZD-202302), the Basic Research Project of ISCAS (Grant No. ISCAS-JCZD-202403), the Youth Innovation Promotion Association of the Chinese Academy of Sciences (Grant Nos. Y2022044 and 2023121), and the Fundamental Research Funds for the Central Universities (Grant No. JK2024-28).
\end{acks}

\bibliographystyle{ACM-Reference-Format}
\bibliography{tosem}


\end{document}